\documentclass[%
 prb,
 reprint,
 nofootinbib,
 amsmath,amssymb,
 aps,
 superscriptaddress,
]{revtex4-2}

\usepackage{graphicx}
\usepackage{caption}
\usepackage{subcaption}
\usepackage{amsmath}
\usepackage{amssymb}
\usepackage{bbold}
\usepackage{comment}
\usepackage{braket}
\usepackage{multirow}
\usepackage{appendix}
\usepackage{wasysym}
\usepackage{physics}
\usepackage{fancyhdr}
\usepackage{}
\usepackage{hyperref}
\hypersetup{
    colorlinks=false,
    linktoc=all,
}

\usepackage{tikz}
\usetikzlibrary{calc}
\newcommand{\tikzmark}[1]{\tikz[overlay,remember picture] \node (#1) {};}
\newcommand{\DrawBox}[4][]{%
    \tikz[overlay,remember picture]{%
        \coordinate (TopLeft)     at ($(#2)+(-0.2em,0.9em)$);
        \coordinate (BottomRight) at ($(#3)+(0.2em,-0.3em)$);
        \path (TopLeft); \pgfgetlastxy{\XCoord}{\IgnoreCoord};
        \path (BottomRight); \pgfgetlastxy{\IgnoreCoord}{\YCoord};
        \coordinate (LabelPoint) at ($(\XCoord,\YCoord)!0.5!(BottomRight)$);
        \draw [#1, color=gray, line width=0bp, rounded corners, fill=gray, fill opacity=0.3] (TopLeft) rectangle (BottomRight);
    }
}

\usepackage[compat=1.1.0]{tikz-feynman}
\usetikzlibrary{decorations.pathmorphing}
\tikzset{snake it/.style={decorate, decoration=snake}}
\definecolor{codegreen}{rgb}{0,0.6,0}
\definecolor{codegray}{rgb}{0.5,0.5,0.5}
\definecolor{codepurple}{rgb}{0.58,0,0.82}
\definecolor{backcolour}{rgb}{0.95,0.95,0.92}

\usepackage{orcidlink}

\newcommand\scalemath[2]{\scalebox{#1}{\mbox{\ensuremath{\displaystyle #2}}}}

\newcommand{\mc}{\mathcal}
\renewcommand{\l}{\left}
\renewcommand{\r}{\right}
\newcommand{\bd}{\boldsymbol}

\newcommand{\dg}{\dagger}
\renewcommand{\u}{\uparrow}
\renewcommand{\d}{\downarrow}
\newcommand{\g}{\hexagon}
\renewcommand{\k}{\davidsstar}

\definecolor{light-gray}{gray}{0.85}
\newcommand{\highlight}[2][light-gray]{\mathchoice%
  {\colorbox{#1}{$\displaystyle#2$}}%
  {\colorbox{#1}{$\textstyle#2$}}%
  {\colorbox{#1}{$\scriptstyle#2$}}%
  {\colorbox{#1}{$\scriptscriptstyle#2$}}}%

\begin{document}
\title{Classification of mass terms in kagome semimetals}
\date{\today}
\author{Simone Ciceri \orcidlink{0009-0003-2088-0156}}
\affiliation{
Institute for Theoretical Physics and Center for Extreme Matter and Emergent Phenomena,
Utrecht University, Princetonplein 5, 3584 CC Utrecht, The Netherlands
}
\affiliation{Einstein Center for Neurosciences Berlin, Charité Universitätsmedizin Berlin, 10117 Berlin, Germany}
\affiliation{Modelling of Cognitive Processes, Technical University of Berlin, 10587, Berlin, Germany}
\author{Matteo Massaro
\orcidlink{0009-0004-0935-8022}}

\affiliation{
Department of Physics and Materials Science\unskip, University of Luxembourg \unskip, L-1511  Luxembourg, Luxembourg
}
\affiliation{
Institute for Theoretical Physics and Center for Extreme Matter and Emergent Phenomena,
Utrecht University, Princetonplein 5, 3584 CC Utrecht, The Netherlands
}
\author{Lars Fritz}
\email[]{l.fritz@uu.nl}
\affiliation{
Institute for Theoretical Physics and Center for Extreme Matter and Emergent Phenomena,
Utrecht University, Princetonplein 5, 3584 CC Utrecht, The Netherlands
}

\begin{abstract}
In the last years, kagome materials received massive attention 
by virtue of being candidate hosts for a large variety of quantum phases: spin liquids, unconventional superconductivity, and topological phases of matter, to name the more exotic.
One of the most interesting features is  tunability: changing the filling, the non-interacting band structure can be tuned from flat bands to conventional metallic phases as well as to semimetals. In this paper we concentrate on the latter. At specific lattice filling the electronic bands have a semimetallic structure, hosting Dirac, massless quasiparticles, like in graphene or other layered two dimensional materials.
Specifically, we determine what terms can be added to the nearest neighbor hopping that open at gap at said Dirac point. These terms can in principle arise through external perturbations, interactions or collective instabilities. We classify the sixteen possible gap-opening terms according to the broken symmetries. Furthermore, we identify concrete microscopic realisations allowing for an interpretation of these phases.
\end{abstract}
\maketitle

\section{\label{sec:intro}Introduction}
Kagome metals have been a focal point for research into two dimensional electronic materials over the last couple of years. Due to the non-trivial geometry of the underlying lattice their non-interacting band structure hosts a variety of interesting configurations, depending on the electronic filling. It hosts conventional metals, a completely flat band, as well as a semimetal like graphene. The flat bands as well as van-Hove singularities makes those systems an interesting playground for the study of strong correlation effects\cite{kang2020topological}.
In addition, they have been proposed as hosts for quantum spin liquids \cite{norman2016colloquium, savary2016quantum}, unconventional superconductivity\cite{mielke2021nodeless}, topological phases of matters\cite{guo2009topological}, as well as other exotic quantum phases\cite{pollmann2014interplay}.

In this article we concentrate on the least studied case of the system which is when the filling is tuned to the semimetallic limit\cite{armitage2018weyl, mazin2014theoretical}.
There, an effective Dirac theory emerges described by massless Dirac fermions. 
We study and classify perturbations to the non-interacting nearest neighbor hopping model, that open a gap at said Dirac points, rendering the quasiparticles massive and effectively turning the system insulating.
This classification has been carried out in graphene a decade ago. It is organized according to symmetries\cite{Chamon_2012} and it analyzes the properties of the underlying insulator\cite{park2015band, semenoff1984condensed, Haldane, kane2005quantum, chamon2000solitons, hou2007electron, yang2019hierarchical, gamayun2018valley, bose2005spin, yazyev2010emergence, li2021spin}.
In this work we carry out a related analysis for kagome and give both a physical interpretation of the mass terms as well as a recipe on how to trigger the respective phases (at least in some cases). 
For some mass terms, the best candidate realisation is a collective instability. Graphene is usually considered a system that is on the verge of being strongly interacting. 
However, Mazin et al. \cite{mazin2014theoretical} showed that Sc-Herbertsmithite is strongly interacting compared to graphene, with a three times larger fine-structure constant\footnote{
The fine-structure constant, defined as
$\alpha = \frac{e^2}{4\pi \epsilon_0 \epsilon_r} \frac{1}{\hbar v_F}$, controls the strength of the Coulomb interaction: $\epsilon_r$ is the relative permittivity of the medium, $e$ the elementary charge and $v_F$ the Fermi velocity of the electrons.}.
This suggests an increased likelihood to undergo spontaneous symmetry breaking to gapped phases.

Our paper is structured as follows.
We initially review the theory of non-interacting, spinless electrons hopping on the kagome lattice.
We then discuss lattice symmetries as well as fundamental symmetries of the underlying quantum theory.
We then classify possible mass terms according to the broken symmetries, allowing reverse-engineer microscopic lattice realizations of perturbations that have said effect.
In parallel, we investigate the relationship with graphene, which shares the same low-energy theory and the underlying lattice symmetries, albeit with a simpler unit cell.

\section{\label{sec:model}Model}
In this section we investigate the band structure of a spinless, non-interacting nearest-neighbor hopping model. From there, we proceed to a derivation of the low-energy theory and a discussion of all its symmetries. 

\subsection{Tight-binding model on the kagome lattice}
\begin{figure}[htp]
    \centering
    \includegraphics[width=\linewidth]{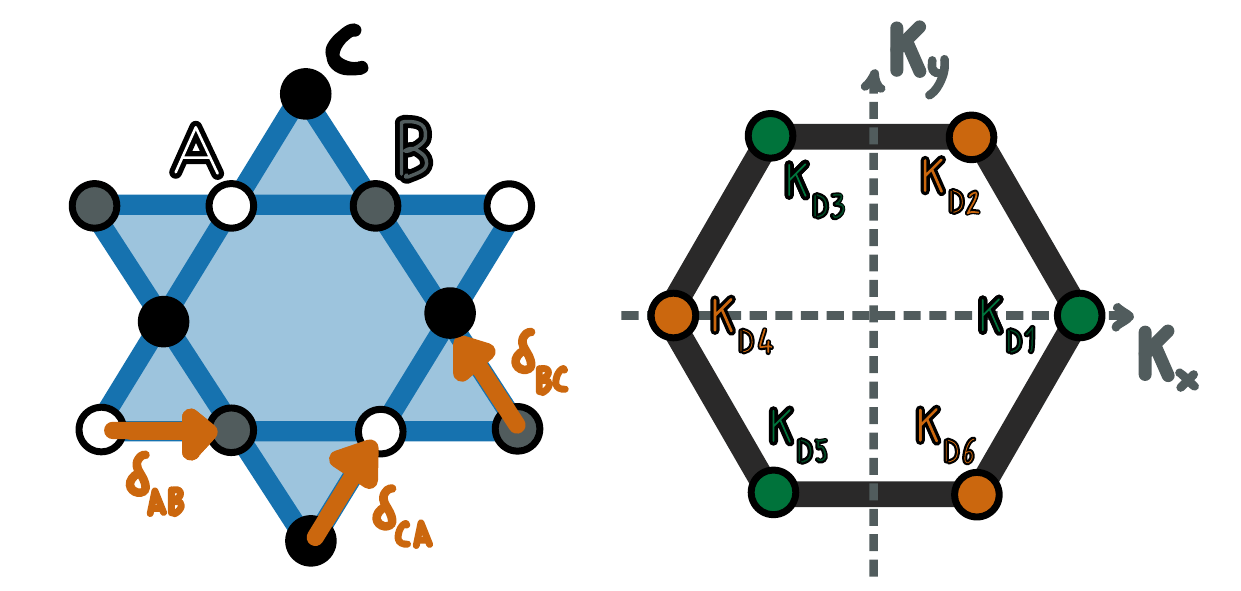}
    \caption{\small Kagome lattice (left figure) and its Brullouin zone (right figure). Sublattices $A, B, C$, sublattice vectors $\delta_{ij}$ and the Dirac points $K_{Dj}$ are indicated.}
    \label{fig:kagomelattice}
    \label{fig:kagomeBZ}
\end{figure}
As a minimal model for the kagome metal, we introduce the tight-binding Hamiltonian for fermions hopping on the kagome lattice. Initially, we disregard spin-orbit coupling and start from spinless particles:
\begin{align}
    H =  -t\sum_{\l<i,j\r>} \psi^{\dagger}_{i} \psi_{j}\;.
    \label{eq:tighbindingH}
\end{align}
The summation extends over nearest-neighbors $\l<i,j\r>$.
The lattice has three sub-lattices, henceforth called $A,B,C$, and lattice vectors 
$
    \boldsymbol{\delta}_{AB} = a(1,0)^T
$, $
    \boldsymbol{\delta}_{BC} = \frac{a}{2}(1, \sqrt{3})^T
$, $
    \boldsymbol{\delta}_{CD} = \frac{a}{2}(-1, \sqrt{3})^T
$, were $a$ is the lattice constant, see Fig.\ref{fig:kagomelattice}.
The Hamiltonian can be block-diagonalized in momentum space with a Fourier transformation, thanks to the discrete translational symmetry of the lattice:
\begin{align}
    H = 
    \sum_{\boldsymbol{k}\in \mathcal{B}}
    \Psi^{\dagger}( \boldsymbol{k})
    h(\boldsymbol{k})
    \Psi( \boldsymbol{k} )\;.
\end{align}
We defined a vector of field operators according to 
\\$\Psi( \boldsymbol{k}) = \Big( \hat{A}(\boldsymbol{k}),   \hat{B}(\boldsymbol{k}), \hat{C}(\boldsymbol{k}) \Big)^T$, where the components refer to the field operators on the respective sublattice.
The Bloch Hamiltonian reads
\begin{align}
    h( \boldsymbol{k} ) = -2 t
    \begin{pmatrix}
    0 & \cos\l( \bd{k} \cdot \boldsymbol{\delta}_{AB}\r) 
        & \cos\l(\boldsymbol{k} \cdot \boldsymbol{\delta}_{AC}\r)
    \\
    \cos\l(\bd{k} \cdot \bd{\delta}_{AB}\r) & 0 & \cos\l( \bd{k} \cdot \bd{\delta}_{BC}\r)
    \\
    \cos\l(\bd{k} \cdot\bd{\delta}_{AC}\r) & \cos\l(\bd{k} \cdot\bd{\delta}_{BC}\r) & 0
    \end{pmatrix}
    \label{eq:BlochHamiltonian}
\end{align}
$\bd{k}=(k_x, k_y)^T$ is a momentum in the first Brillouin zone shown in Fig.~\ref{fig:kagomeBZ}.
A diagonalization of the Bloch Hamiltonian reveals three bands:
\begin{align}
    E(\bd{k}) =  
    \begin{cases}
    -t \Big( 1 \pm \sqrt{3 + 2\sum_{i<j} \cos\l( 2 \bd{k} \cdot \bd{\delta}_{ij} \r) } \Big)
    \\
    2 t
    \end{cases}\;.
\end{align}
The two lower bands touch at six Dirac points, among which only two are independent. We refer to those as \textbf{valleys}. A possible choice is $\bd{K}_{D1} = \l(\frac{2\pi}{3 a}, 0\r)^T$ and $\bd{K}_{D4} = \l(-\frac{2\pi}{3 a}, 0\r)^T$, pointed in Figure \ref{fig:kagomeBZ}. Different choices of Dirac points result in different low energy Hamiltonians that are related by gauge transformations (see discussion in Appendix \ref{sec:appendix}).

\begin{figure}[htp]
        \centering
        \includegraphics[width=\linewidth]{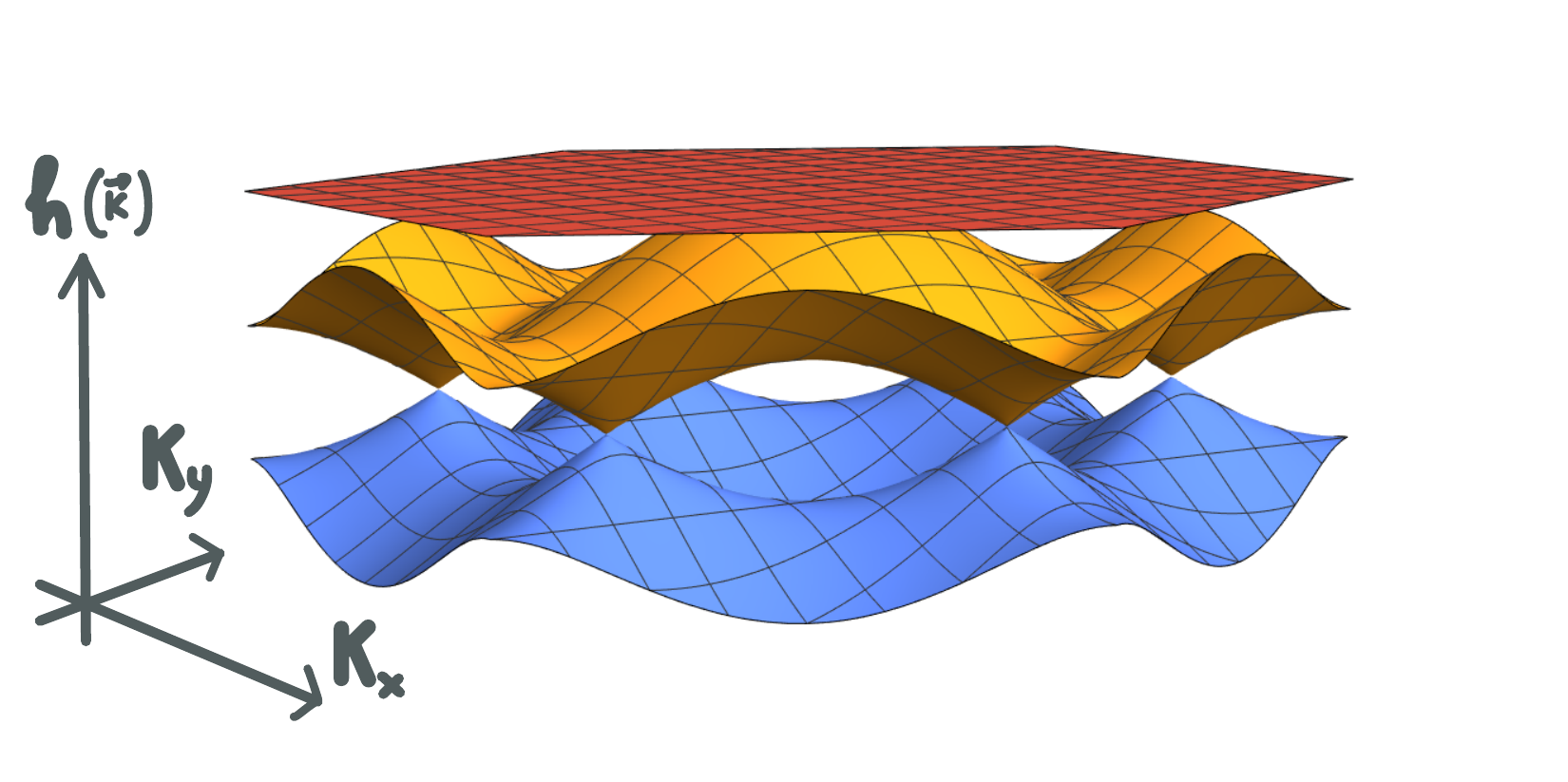}
        \caption{\small Electronic bands of the tight-binding model on the kagome lattice: a flat band (red), and two lower bands (orange, blue) that touch at six Dirac points where the density of states vanishes. The focus of this work is on the low energy theory that emerges when the filling is tuned at these Dirac points (for instance, when only the lower, blue band is completely filled).} 
    \end{figure}

\subsubsection{Low energy expansion}
In a first step we want to investigate the mass terms in the low-energy theory. To that end, we will expand the Block Hamiltonian to linear order in $\bd{k}$ and perform a unitary transformation.
This reveals the two-dimensional effective Dirac theory. The diagonalization matrix is given by:
\begin{align}
    U = 
    \begin{bmatrix}
    \frac{1}{\sqrt{2}} 
        \begin{pmatrix}
        0 \\ 1 \\ 1
        \end{pmatrix}
    ,&
    \frac{1}{\sqrt{6}}
    \begin{pmatrix}
        2 \\ -1 \\ 1
        \end{pmatrix}
    ,&
    \frac{1}{\sqrt{3}}
    \begin{pmatrix}
        -1 \\ -1 \\ 1
    \end{pmatrix}
    \end{bmatrix} 
    \label{eq:kagomedecoupledfields}
\end{align}
We consider the expansion of the transformed Hamiltonian $\Tilde{h}(\bd{k}) = U h(\bd{K}_{Di}+\bd{k}) U^{-1}$ up to linear order in $\bd{k}$, centered at the two Dirac points $\bd{K}_{D1}$ and $\bd{K}_{D4}$.
In terms of two new momenta 
$
    k_1 = \frac{1}{2}(k_x - \sqrt{3} k_y)
$ and $
    k_2 = \frac{1}{2}(\sqrt{3}k_x + k_y)
$,
the transformed, linearized Hamiltonian reads
\begin{align}
    &\Tilde{h}(\bd{K}_{D1}+\bd{k}) \simeq \begin{pmatrix}
        \tikzmark{left1} 
        -t-v_f k_1 & -v_f k_2 & \frac{v_f}{\sqrt{2}} k_2
        \\
        -v_f k_2 & -t+v_f k_1  \tikzmark{right1}
        & \frac{v_f}{\sqrt{2}} k_1
        \\
        \frac{v_f}{\sqrt{2}} k_2 & \frac{v_f}{\sqrt{2}} k_1 & 2t
    \end{pmatrix}
    \label{eq:HD}
    \\[0.2cm]
    &\Tilde{h}(\bd{K}_{D4}+\bd{k}) \simeq \Tilde{h}(\bd{K}_{D1}-\bd{k})
    \label{eq:HD2}
\end{align}
\DrawBox[thick, red ]{left1}{right1}{\textcolor{red}}
where $v_f = \sqrt{3} a t$ can be seen as the effective Fermi velocity. Note that the Jacobian of this transformation is one.
The highlighted 2x2 block in the left-upper corner describes the emerging Dirac theory, while the other terms couple it to the flat band with $E=2t$.
Integrating out this flat band contributes terms of order $\mathcal{O}(\bd{k}^2)$ to the Dirac theory and can be therefore neglected.
For a more detailed description of these calculations, we refer to a previous work\cite{ciola2021chiral} where this is shown in more detail.

In the two-dimensional subspace of the Dirac theory, the Hamiltonian reads
\begin{align}
    & \Bar{h}(\bd{K}_{D1} + \bd{k}) = -t \mathbb{1} - v_f \l( k_1 \hspace{0.1cm} \tau_3 + k_2 \hspace{0.1cm} \tau_1  \r),
    \\
    & \Bar{h}(\bd{K}_{D4} + \bd{k}) = \Bar{h}(\bd{K}_{D1} - \bd{k}),
\end{align}
where $\tau_i$ is the $i$-th Pauli matrix. Note that the corresponding field operators, defined through Eq.\ref{eq:kagomedecoupledfields}, are linear combinations of the three sublattices fields $ \hat{A},\hat{B},\hat{C}$.
We later refer to this degree of freedom as \textbf{\textit{sublattice}}, in analogy with the sublattices of graphene. This is abuse of language, due to the presence of three sublattices and the wavefunction being distributed over all of them. 
At 1/3 lattice filling, the term proportional to the identity is absorbed in the chemical potential and we can neglect it. A more general form of the Hamiltonian can be written, including the spin and valley degrees of freedom, according to
\begin{align}
     \highlight{
     \Tilde{\mathcal{H}}( \bd{k} ) =  
     - v_f
    \l(k_1 \hspace{0.12cm} s_0 \otimes \sigma_3\otimes \tau_3 
     +  k_2 \hspace{0.12cm} s_0 \otimes \sigma_3\otimes \tau_1
     \r)
     }
     \label{eq:H2dim}
\end{align}
where $s_i$ and $\sigma_i$ are the $i$-th Pauli matrices acting on the spin and valley subspace, respectively. The valley dependence is derived from Eqs.~\eqref{eq:HD},\eqref{eq:HD2}. The identity in the spin space reflects the absence of spin-orbit coupling, that is studied in the next chapter in the context of perturbations.
The field operators in this picture are
\small
\begin{equation}
    \scalemath{0.9}{
    \hat{\Psi}( \bd{k}) = 
    \setlength\arraycolsep{1.5pt}
    \begin{pmatrix}
      \hat{\varphi}_{\bd{k}\u+}, 
    & \hat{\eta}_{\bd{k}\u+}, 
    & \hat{\varphi}_{\bd{k}\u-}, 
    & \hat{\eta}_{\bd{k}\u-}, 
    & \hat{\varphi}_{\bd{k}\d+}, 
    & \hat{\eta}_{\bd{k}\d+}, 
    & \hat{\varphi}_{\bd{k}\d-}, 
    & \hat{\eta}_{\bd{k}\d-} 
    \end{pmatrix}^T
    }
\end{equation}
\normalsize
where $\hat{\varphi}$, $\hat{\eta}$ represent the two \textit{sublattice} field operators with defined valley ($\pm$) 
and spin quantum numbers ($\u \d$).
The respective low-energy effective Hilbert space is given by
\begin{align}
    \mathbb{H}_{\text{eff}} \equiv \underbrace{\mathbb{C}^2}_\text{Spin space}\otimes\underbrace{\mathbb{C}^2}_\text{Valleys sp.}\otimes\underbrace{\mathbb{C}^2}_\text{\textit{Sublatt}. sp.}
\end{align}
For later reference, when we reverse-engineer couplings in the microscopic model where the three sublattices are explicitly represented, we refer to the Hilbert space
\begin{align}
    \mathbb{H}_{\text{full}} \equiv \underbrace{\mathbb{C}^2}_\text{Spin space}\otimes\underbrace{\mathbb{C}^2}_\text{Valleys sp.}\otimes\underbrace{\mathbb{C}^3}_\text{Sublatt. sp.}
\end{align}
which is obviously higher dimensional.

\subsection{Symmetries of the model}
\begin{figure}[htp]
    \centering
    \includegraphics[width=\linewidth]{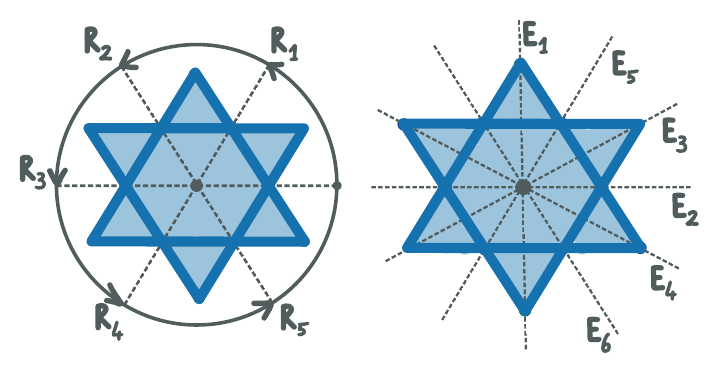}
    \caption{
    \small Rotation symmetries $R_{i}$ (left figure) and reflection symmetries $E_{i}$ (right figure) of the kagome lattice. The tight-binding Hamiltonian Eqs. \eqref{eq:tighbindingH},\eqref{eq:BlochHamiltonian} is invariant under these transformations. However, Dirac points are interchanged by rotations and reflections. A linear expansion of the Hamiltonian around two specific Dirac points is accordingly transformed into a linear expansion around two different ones. Explicit examples are shown in Appendix \ref{sec:appendix}).
    }
    \label{fig:reflections}
\end{figure}

The lattice symmetries (and equivalently those of the honeycomb lattice) are encoded in the wallpaper group \textit{p6m}: it consists of six reflection axes, one rotation center of order six, two of order three, and three of order two \cite{flores2007classifying}.
Reflections and the six rotations with the same center are shown in  Fig.~\ref{fig:reflections}. Their operator form, which commutes with the tight-binding Hamiltonian, is derived explicitly in Appendix ~\ref{sec:appendix}. The main results are summarized in Table~\ref{tab:symmetry_operators}. 
Note that rotations and reflections interchange Dirac points. 
$R_1$ for instance, the anti-clockwise rotation of $\pi/3$, transforms $K_{D1}$ and $K_{D4}$ into $K_{D2}$ and $K_{D5}$ respectively. A linear expansion of the Hamiltonian around two specific Dirac points is consequently transformed into the linear expansion around two corresponding ones.

Besides lattice symmetries and discrete translational symmetry (accounted for with the Fourier transform), we consider three additional symmetries: time reversal, chirality, and charge conjugation.
In the low-energy effective Hilbert space $\mathbb{H}_{\text{eff}}$, time-reversal assumes the form
$
    \mathcal{T} = U_T \cdot \mc{K} = 
    i s_2 \otimes \sigma_1 \otimes \tau_0 \cdot \mc{K}
$
where $\mc{K}$ represents complex conjugation and $U_T$ is unitary. 
The Hamiltonian is $\mathcal{T}$-invariant since $U_T \tilde{\mathcal{H}}^*({\bd{k}})U_T^{-1} = \tilde{\mathcal{H}}(-{\bd{k}})$.
Charge conjugation is an anti-unitary symmetry that relates a particle excitation to a hole excitation of opposite charge and opposite energy.
Chirality (also known as sublattice symmetry), is defined as the composition of time-reversal and charge symmetry, and it is unitary.
In contrast to graphene, the kagome tight-binding Hamiltonian enjoys neither chiral symmetry, nor charge symmetry. In simple terms, it is due to the flat band which makes the spectrum asymmetric under a sign flip $t \to -t$. 
However, the low-energy theory Eq.\ref{eq:H2dim} obtained by integrating out the flat band is charge and chiral symmetric.
The chirality operator
$
    \mathcal{S} = s_0 \otimes \sigma_3 \otimes \tau_2
$ anticommutes with the effective Dirac Hamiltonian: $\l\{ \tilde{\mathcal{H}}({\bf{k}}), \mc{S} \r\}=0$.
Charge conjugation $\mathcal{C}$ is obtained by the combination of time-reversal and the inverse of sublattice symmetry:
$
    \mc{C} = \mc{T} \cdot \mc{S}^{-1}
$.
The simplest form is obtained in the low-energy subspace $\mathbb{H}_{\text{eff}}$, where
$
    \mc{C} = U_{\mc{C}} \cdot \mc{K} =  s_2 \otimes \sigma_2 \otimes \tau_2 \cdot \mc{K}
$.
The effective Hamiltonian is charge-symmetric since $U_{\mc{C}} \tilde{\mathcal{H}}^*(-{\bf{k}}) U_{\mc{C}}^{-1} = -\tilde{\mathcal{H}}({\bf{k}}) $.
Symmetry relations for $\mathcal{C}$ and $\mc{S}$ only hold if the flat band is neglected. This approximation is justified when the theory is expanded around the Dirac cones and only low-energy effects are retained, as illustrated in the previous section.
To summarize, while the tight-binding theory of graphene is charge and chiral-symmetric, this holds in kagome only when the flat band is ignored and the filling is tuned to the Dirac point.

\section{\label{sec:massterms} Mass terms}
There are many different bilinear terms that could be added to the band structure introduced above. We concentrate on bilinear terms that open a gap at the Dirac points: these terms convert the semimetal to an insulator. 
The most general local bilinear term involves spin and valley degrees and freedom, and within the low-energy subspace $\mathbb{H}_{\text{eff}}$ it takes the form
$
    m 
    \mc{M}_{ijk} 
    = m \hspace{0.11cm} 
    s_i \otimes \sigma_j \otimes \tau_k
$.
The scalar $m$ encodes the amplitude and $s_i$, $\sigma_j$ and $\tau_k$ are Pauli matrices acting on the spin, valley and \textit{sublattice} subspaces, respectively.
If $\mc{M}_{ijk}$ anti-commutes with the Dirac Hamiltonian, {\it i.e.},
$\l\{\tilde{\mc{H}}, \mc{M} \r\}=\tilde{\mc{H}}\mc{M} + \mc{M}\tilde{\mc{H}} =  0$, this results in a gap in the excitation spectrum.
The resulting dispersion relation reads:
$
    \epsilon_{\bd{k}} = \pm \sqrt{\bd{k}^2 + m^2}
    \label{eq:dispersion}
$.
From now on, we refer to all anti-commuting terms as \textbf{gap terms} or \textbf{mass terms}.
According to our classification, these operators do not involve couplings to the flat band. We embed the gap terms in the full space according to
\begin{align}\label{embed_mass_term_in_H_full}
    \Tilde{M}_{ijk} = s_i \otimes \sigma_j \otimes
    \begin{pmatrix}
        \multicolumn{2}{c}{\multirow{2}{*}{\Large $\tau_k$}} & 0\\
        & & 0\\
        0 & 0 & 0 \\
    \end{pmatrix}\;.
\end{align}
While there are 64 possible bilinear terms, only 16, reported in Tab.\ref{tab:kagome_mass_resume}, anti-commute with the Hamiltonian and represent mass terms. 
Note that $\Tilde{M}_{ijk}$ is written in the decoupled representations, in which the Hamiltonian is diagonal at the Dirac points.
The mass terms in the original representation of the full Hilbert-space (same as Eq. \ref{eq:BlochHamiltonian}) are obtained using
$
    M_{ijk} = 
    U^{-1}
    \Tilde{M}_{ijk}
    U
    \label{eq:massTerm}
$.
We now proceed to classify the mass terms according to the broken and preserved symmetries (Appendix \ref{sec:appendix}).
An analogous investigation can be found for graphene in Ref.~\cite{Chamon_2012}.

\section{\label{sec:latticerealisations} Microscopic realisation of mass terms}
\begin{table*}
    \centering
    \begin{tabular}{|c||c|c|c|c||c|c|}
    \hline
    \multicolumn{7}{c}{\textbf{Kagome mass terms $\mc{M}_{ijk}$}} \\
    \hline        
Idx $\{ijk\}$ & Parity & T-rev & Char & Chiral & Lattice realisation & Section \\ 
\hline \hline
 \{0,0,2\} & $\color{green}{\checkmark}$ & $\color{red}{\times}$ & $\color{green}{\checkmark}$ & $\color{red}{\times}$
 & Haldane's model (spin-chiral) & \ref{sec:HaldaneKagome}
 \\
 \{0,1,0\} & $\color{green}{\checkmark}$ & $\color{green}{\checkmark}$ & $\color{green}{\checkmark}$ & $\color{green}{\checkmark}$
 & Plaquette phase (rings) & \ref{sec:ResPlaquetteKagome}
 \\
 \{0,2,0\} & $\color{red}{\times}$ & $\color{green}{\checkmark}$ & $\color{green}{\checkmark}$ & $\color{green}{\checkmark}$
  & Plaquette phase (rings) & \ref{sec:ResPlaquetteKagome}
 \\
 \{0,3,2\} & $\color{red}{\times}$ & $\color{green}{\checkmark}$ & $\color{red}{\times}$ & $\color{red}{\times}$ 
 & Lattice dimerization \& trimerization & \ref{sec:dimerization_kagome}
 \\
 \hline
 \{1,1,0\} & $\color{green}{\checkmark}$ & $\color{red}{\times}$ & $\color{red}{\times}$ & $\color{green}{\checkmark}$ 
 & Resonating plaquette phase - $x$ & \ref{sec:ResPlaquetteKagome}
 \\
 \{2,1,0\} & $\color{green}{\checkmark}$ & $\color{red}{\times}$ & $\color{red}{\times}$ & $\color{green}{\checkmark}$
 & Resonating plaquette phase - $y$ & 
 \\
 \{3,1,0\} & $\color{green}{\checkmark}$ & $\color{red}{\times}$ & $\color{red}{\times}$ & $\color{green}{\checkmark}$ 
 & Resonating plaquette phase - $z$ & 
 \\
 \{1,2,0\} & $\color{red}{\times}$ & $\color{red}{\times}$ & $\color{red}{\times}$ & $\color{green}{\checkmark}$ & Resonating plaquette phase - $x$ &
 \\
 \{2,2,0\} & $\color{red}{\times}$ & $\color{red}{\times}$ & $\color{red}{\times}$ & $\color{green}{\checkmark}$ & Resonating plaquette phase - $y$ & \\
 \{3,2,0\} & $\color{red}{\times}$ & $\color{red}{\times}$ & $\color{red}
 {\times}$ & $\color{green}{\checkmark}$ & Resonating plaquette phase - $z$ & \\
\hline
 \{1,0,2\} & $\color{green}{\checkmark}$ & $\color{green}{\checkmark}$ & $\color{red}{\times}$ & $\color{red}{\times}$ & Spin-orbit coupling - $x$ & \ref{sec:SOC_kagome} \\
 \{2,0,2\} & $\color{green}{\checkmark}$ & $\color{green}{\checkmark}$ & $\color{red}{\times}$ & $\color{red}{\times}$ & Spin-orbit coupling - $y$ & \\
 \{3,0,2\} & $\color{green}{\checkmark}$ & $\color{green}{\checkmark}$ & $\color{red}{\times}$ & $\color{red}{\times}$ & Spin-orbit coupling - $z$ & \\
 \hline
 \{1,3,2\} & $\color{red}{\times}$ & $\color{red}{\times}$ & $\color{green}{\checkmark}$ & $\color{red}{\times}$ & SOC in trimerized latt. $x$ & \ref{sec:SOCtrimerizedKagome}\\
 \{2,3,2\} & $\color{red}{\times}$ & $\color{red}{\times}$ & $\color{green}{\checkmark}$ & $\color{red}{\times}$ & SOC in trimerized latt. $y$ & \\
 \{3,3,2\} & $\color{red}{\times}$ & $\color{red}{\times}$ & $\color{green}{\checkmark}$ & $\color{red}{\times}$ & SOC in trimerized latt. $z$ & \\
 \hline
    \end{tabular}
    \caption{Kagome mass terms and their physical interpretation. Indices $i,j,k$ in the first column label each term according to 
    $
    m 
    \mc{M}_{ijk} 
    = m \hspace{0.11cm} 
    s_i \otimes \sigma_j \otimes \tau_k.
    $
   The quantum symmetries of the full massless Hamiltonian (parity, time reversal, charge, and chiral symmetry) are either preserved ($\color{green}{\checkmark}$) or broken ($\color{red}{\times}$) by each term.
 }
    \label{tab:kagome_mass_resume}
\end{table*}

The main result of our study and focus of this section is to present lattice realisations for each of the possible mass terms. We start every paragraph with a table that lists all the symmetries. In order to distinguish their presence or absence, we note “preserved" with a green checkmark ($\color{green}{\checkmark}$) while we denote “broken" with a red cross ($\color{red}{\times}$). Time-reversal symmetry is denoted $\mc{T}$, charge conjugation $\mc{C}$, and chirality $\mc{S}$, while reflections are written as $E_i$ and rotations as $R_{i}$.
Table \ref{tab:kagome_mass_resume} summarises the following discussion.

\subsection{Haldane's model on kagome lattice}\label{sec:HaldaneKagome}
\begin{figure}[htp]
    \includegraphics[width=\linewidth]{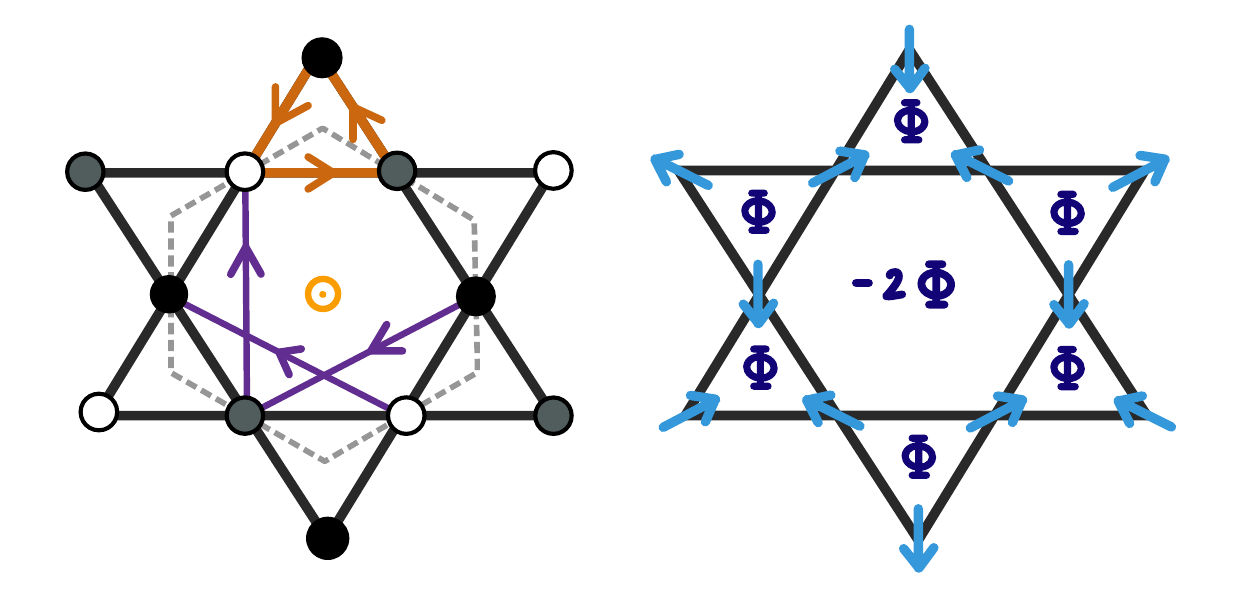}
    \caption{\small
    \textbf{(Left figure)} Kagome unit cell, with nearest neighbour (NN, orange) and second-nearest neighbour (SNN, purple) bonds. Arrows mark the direction of positive phase hoppings when Haldane's periodic magnetic field is present.
    A dotted, grey line encloses a surface of vanishing magnetic flux.
    A magnetic field of this kind results in the mass term $\mc{M}_{002}$ that opens a gap at the Dirac points.
    \textbf{(Right figure)}
    Magnetic fluxes $\Phi$ through the triangles and the hexagons when Haldane's magnetic field is present. In addition, we show the spin-chiral arrangement on the kagome lattice background. When electron spins are perfectly aligned to the nuclei spins (strong Hund's coupling), this chiral background order effectively acts as the Haldane's magnetic field.
    }
    \label{fig:haldane_kagome}
\end{figure}

\begin{center}
\tabcolsep=0.06cm
\small
\begin{tabular}{|c|c|c|c|c|c|c|c|c|c|c|c|c|c|c|}
\hline
\multicolumn{1}{|c|}{\multirow{2}{*}{ $\mc{M}_{002}$}}  & $\mc{T}$ & $\mc{C}$ & $\mc{S}$ & E1 & E2 & E3 & E4 & E5 & E6 & R1 & R2 & R3 & R4 & R5
\\
\cline{2-15}
\multicolumn{1}{|c|}{} & $\color{red}\times$ & $\color{green}\checkmark$ & $\color{red}\times$ & $\color{red}\times$ & $\color{red}\times$ & $\color{red}\times$ & $\color{red}\times$ & $\color{red}\times$ &
   $\color{red}\times$ & $\color{green}\checkmark$ & $\color{green}\checkmark$ & $\color{green}\checkmark$ & $\color{green}\checkmark$ & $\color{green}\checkmark$
   \\
\hline
\end{tabular}
\end{center}

In this section, we explore the effect of a spatially dependent magnetic field $\bd{\mc{B}}(\bd{r}) = \mc{B}(\mathbf{r}) \mathbf{z}$ perpendicular to the 2D plane. It is chosen such that there is zero flux through the unit cell. 
Following the path of Haldane's work in graphene\cite{Haldane}, we consider a magnetic field with the periodicity of the lattice. For instance, this configuration can be obtained by placing magnetic dipoles at the centre of each hexagon\cite{liu2010simulating}. 
Using Peierls substitutions we absorb the effect of the magnetic field into the hopping amplitude $t$, which acquires a complex phase:
$t \to t e^{i\pm\theta}$. 
The kagome structure causes finite fluxes of $\bd{\mc{B}}$ through closed paths of nearest neighbour hoppings.
The phase $\theta$ is therefore finite for NN paths, as opposed to graphene where next-nearest neighbours (NNN) hoppings give the first, non-zero contribution.
Given $\Phi$ as the flux of the magnetic field through the triangles, the phase is computed using the Stokes theorem:
$
    \theta = \frac{e}{3\hbar} \Phi
$.
The sign of the acquired phase depends on the path direction, as shown in Figure \ref{fig:haldane_kagome}: hoppings as $A\to B$, $B\to C$, $C\to A$ will gain a phase $+\theta$, while a phase $-\theta$ is obtained for opposite hoppings.
The TB Hamiltonian between nearest neighbours becomes 
\begin{equation}
\begin{split}
    H_\theta = 
    t \sum_{i}
    e^{-i\theta}
    \Big[ 
    \hat{A}^\dg_i \hat{B}_{i\pm\bd{\delta}_{AB}} 
    & + 
    \hat{B}^\dg_{i+\bd{\delta}_{AB}} \hat{C}_{i\pm\bd{\delta}_{AC}} 
    \\
    & 
    + \hat{C}^\dg_{i\pm\bd{\delta}_{AC}}\hat{A}_{i} 
    \Big] + h.c.
\end{split}
\end{equation}
where $h.c.$ indicates the hermitian conjugate.
After the Fourier transform, we obtain
\begin{align}
    h_\theta(\bd{k}) =
    \everymath={\scriptstyle}
    \begin{pmatrix}
    0 & e^{-i\theta}\cos\l(\bd{k} \cdot \bd{\delta}_{AB}\r) & e^{i\theta}\cos\l( \bd{k} \cdot \bd{\delta}_{AC}\r) 
    \\
    e^{i\theta}\cos\l(\bd{k} \cdot \bd{\delta}_{AB}\r)  & 0 & e^{-i\theta}\cos\l(\bd{k}\cdot\bd{\delta}_{BC}\r) 
    \\
    e^{-i\theta}\cos\l( \bd{k} \cdot \bd{\delta}_{AC}\r)   &   e^{i\theta}\cos\l(\bd{k}\cdot\bd{\delta}_{BC}\r) & 0
    \end{pmatrix}
\end{align}
We expand the Hamiltonian to linear order around the two valleys and include the spin degree of freedom. The resulting Dirac Hamiltonian reads
\begin{equation}
\begin{split}
    h_\theta
    = 
    &t \cos \theta 
    \l[
    s_0
    \otimes
    \sigma_0
    \otimes
    \begin{pmatrix}
    0 & -1 & 1\\
    -1 & 0 & 1\\
    1 & 1 & 0 \\
    \end{pmatrix}
    \r]
    + 
    \\
    &-  t \sin \theta
    \l[
    s_0
    \otimes
    \sigma_0
    \otimes
    \begin{pmatrix}
    0 & -i & -i\\
    i & 0 & i\\
    i & -i & 0 \\
    \end{pmatrix}
    \r]
    \label{eq:haldaneKagomeNN}
\end{split}
\end{equation}
The first term is equal to the TB Hamiltonian in the absence of a magnetic field, multiplied by a factor $\cos\theta$. The second term instead corresponds to $\mc{M}_{002}$, which opens a gap at the Dirac points. 
The strength of this perturbation is $-t\sin\theta$. Since $\theta$ is proportional to the flux of $\bd{\mc{B}}(\mathbf{r})$, the Hamiltonian \ref{eq:haldaneKagomeNN} reduces to the original TB form when $\mc{B}(\mathbf{r})=0$. Otherwise, a gap of size $\Delta = \l| t\sqrt{3} \sin \theta \r|$ opens at the Dirac points when the magnetic field is switched on. 
If NNN hoppings are taken into account the gap is further increased.
Haldane's mass term breaks time-reversal and chiral symmetry, keeping charge symmetry intact. From a geometrical perspective, all the reflection symmetries are broken, while all the rotations are preserved, as one can check from Figure \ref{fig:haldane_kagome}.

{\it{Alternative realization through Hund's coupling and spin chirality:}}
We considered the emergence of Haldane's mass by artificially applying an external, fine-tuned magnetic field. 
The same effect, however, can originate from the interplay between conduction electrons and a non-trivial magnetic background \cite{nagaosa2010anomalous, taguchi2001spin, ye1999berry}. 
In particular, conduction electrons strongly coupled to the nuclear spins (Hund's coupling) acquire a Berry phase if their spin is forced to align with the background spins:
a configuration of tilted spins in the underlying structure acts as an effective magnetic field \cite{haldane2004berry}.
The kagome lattice promotes one peculiar spin configuration, with a pattern of alternating spin-chirality \cite{kolincio2022kagom}, represented in Figure \ref{fig:haldane_kagome}.
The effective magnetic field resulting from this background configuration is identical to the magnetic field of Haldane's model and is responsible for the same gap-opening in the spectrum.

\subsection{Spin-orbit coupling and the topological insulator}\label{sec:SOC_kagome}
\begin{center}
\tabcolsep=0.06cm
\small
\begin{tabular}{|c|c|c|c|c|c|c|c|c|c|c|c|c|c|c|}
\hline
  & $\mc{T}$ & $\mc{C}$ & $\mc{S}$ & E1 & E2 & E3 & E4 & E5 & E6 & R1 & R2 & R3 & R4 & R5
\\
\hline
$\mc{M}_{102}$ & $\color{green}\checkmark$ & $\color{red}\times$ & $\color{red}\times$ & $\color{red}\times$ & $\color{green}\checkmark$ & $\color{red}\times$ & $\color{red}\times$ & $\color{red}\times$ &
   $\color{red}\times$ & $\color{red}\times$ & $\color{red}\times$ & $\color{red}\times$ & $\color{red}\times$ & $\color{red}\times$ \\
\hline
$\mc{M}_{202}$ & $\color{green}\checkmark$ & $\color{red}\times$ & $\color{red}\times$ & $\color{green}\checkmark$ & $\color{red}\times$ & $\color{red}\times$ & $\color{red}\times$ & $\color{red}\times$ &
   $\color{red}\times$ & $\color{red}\times$ & $\color{red}\times$ & $\color{red}\times$ & $\color{red}\times$ & $\color{red}\times$ \\
\hline
$\mc{M}_{302}$ & $\color{green}\checkmark$ & $\color{red}\times$ & $\color{red}\times$ & $\color{green}\checkmark$ & $\color{green}\checkmark$ & $\color{green}\checkmark$ & $\color{green}\checkmark$ & $\color{green}\checkmark$ & $\color{green}\checkmark$ & $\color{green}\checkmark$
   & $\color{green}\checkmark$ & $\color{green}\checkmark$ & $\color{green}\checkmark$ & $\color{green}\checkmark$ \\
\hline
\end{tabular}
\end{center}

In kagome structures, as opposed to graphene, the intrinsic spin-orbit coupling (SOC) is relevant for nearest-neighbour hopping: NN paths do not correspond to any reflection symmetry axis of the lattice, resulting in finite angular momentum.
The sign of the SOC amplitude $\pm i \lambda$ depends on the spin and hopping direction, as shown in Figure \ref{fig:haldane_kagome}. 
Guo and Franz \cite{guo2009topological} explored a similar interaction, considering instead hoppings between next-nearest neighbors.
We expect however the SOC in NN hops to be dominant.
The SOC Hamiltonian is given by
\begin{align*}
    H_{soc} &= \pm i \frac{\lambda}{2}
    \sum_{\alpha,\beta}
    \sum_{\l< i,j\r>}  \psi^\dg_{i\alpha} \psi_{j\beta}
    \l(s_3\r)_{\alpha\beta}
    \\
    &= 
    \pm i \frac{\lambda}{2} \sum_{\l< i,j\r>}
    \l( 
    \psi^\dg_{i\u} \psi_{j\u}
    -
    \psi^\dg_{i\d} \psi_{j\d}
    \r)
    \label{eq:SOC_kagome}
\end{align*}
where $\pm$ is related to the hopping direction pictured in Figure \ref{fig:haldane_kagome}, and $s_3$ is the third Pauli matrix in spin space.
The explicit form in terms of the sublattice fields is
\begin{equation}
\begin{split}
    H_{soc} = i \frac{\lambda}{2} \sum_{i \in A}  
    \Big( 
    \hat{A}^\dg_{i,\u} \hat{B}_{i\pm \delta_{AB}, \u}
    + 
    \hat{B}^\dg_{i+\delta_{AB},\u} \hat{C}_{i\pm \delta_{AC}, \u}
    + 
    \\
    +
    \hat{C}^\dg_{i\pm \delta_{AC},\u} 
    \hat{A}_{i,\u}
    -
    \boxed{\u 
    \hspace{0.2cm}
    \leftrightarrow
    \hspace{0.2cm}\d}
    \Big)
    +
    h.c.   
\label{eq:SOC_kagome}
\end{split}
\end{equation}
Here, $\boxed{\u 
    \hspace{0.2cm}
    \leftrightarrow
    \hspace{0.2cm}\d}$
indicates the corresponding term with inverted spin indices. 
After Fourier transform, this reads
\begin{align*}
    &H_{soc} = 
    \sum_{\bd{k}}
    \hat{\Phi}^\dg_{\bd{k}}
    h_{soc}(\bd{k})
    \hat{\Phi}_{\bd{k}}
    \\
    &h_{soc}(\bd{k}) = 
    i \lambda s_3 \otimes
    \everymath{\scriptstyle}
    \begin{pmatrix}
    0 & \cos\l(\bd{k}\cdot\bd{\delta}_{AB}\r) & -\cos\l(\bd{k}\cdot\bd{\delta}_{AC}\r)
    \\
    -\cos\l(\bd{k}\cdot\bd{\delta}_{AB}\r) & 0 & \cos\l(\bd{k}\cdot\bd{\delta}_{BC}\r)
    \\
    \cos\l(\bd{k}\cdot\bd{\delta}_{AC}\r) & -\cos\l(\bd{k}\cdot\bd{\delta}_{BC}\r) & 0
    \end{pmatrix}
\end{align*}
Here the field operators are defined as
$\hat{\Phi}_k = 
\begin{pmatrix}
\hat{A}_{k\u}, & \hat{B}_{k\u}, & \hat{C}_{k\u}, & 
\hat{A}_{k\d}, & \hat{B}_{k\d}, & \hat{C}_{k\d} 
\end{pmatrix}^T
$.
Evaluating the Bloch Hamiltonian $h_{soc}(\bd{k})$ at the two Dirac points, and including the valley degree of freedom one obtains
\begin{equation}
    \mc{H}_{soc} = 
    i \frac{\lambda}{2}
    \hspace{0.07cm}
    s_3 \otimes
    \sigma_0 \otimes
    \begin{pmatrix}
    0 & -1 & -1
    \\
    1 & 0 & 1
    \\
    1 & -1 & 0
    \end{pmatrix}
\end{equation}
which corresponds to the gap-opening term $\mc{M}_{302}$.
SO coupling for second-nearest neighbours hopping corresponds to the same mass term \cite{guo2009topological}, with a different SOC amplitude $\lambda'$. Since $\lambda$ and $\lambda'$ have opposite signs, the gap amplitude is reduced as $|\lambda|-|\lambda'|$.
Time-reversibility is preserved, as well as all reflections and rotations.    
No $x$ or $y$ spin couplings are present in the case of planar kagome lattice, because of the $\hat{z} \to -\hat{z}$ reflection symmetry.
If a curvature is present, however, the angular momentum vector acquires non-zero components in the $x$ and $y$ directions, resulting in additional terms proportional to $\mc{M}_{102}$ and $\mc{M}_{202}$.
Such couplings respect time-reversal symmetry but break certain lattice symmetries which were preserved in the planar case.
Guo and Franz demonstrated the presence of gapless edge states in the SO phase, which characterize it as a topological insulator \cite{guo2009topological}.

\subsection{(Resonating) Plaquette ordered phases}\label{sec:ResPlaquetteKagome}
\begin{figure}[htp]
    \centering
    \includegraphics[width=\linewidth]{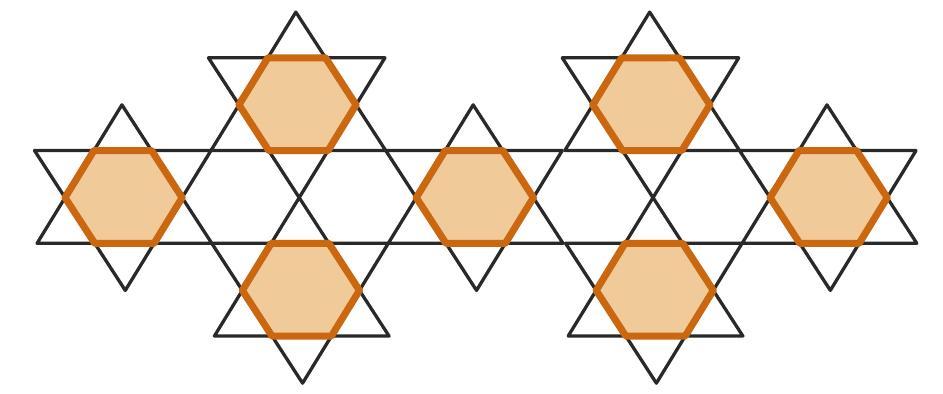}
    \caption{
    \small O-type Kekulé dimerization pattern on kagome lattice, also referred to as plaquette ordered phase. A mean-field approximation of the interaction with this background charge ordering results in the mass terms $\mc{M}_{010}$, $\mc{M}_{020}$.}
    \label{fig:plaquettes_kagome}
\end{figure}

\begin{center}
\tabcolsep=0.06cm
\small
\begin{tabular}{|c|c|c|c|c|c|c|c|c|c|c|c|c|c|c|}
\hline
  & $\mc{T}$ & $\mc{C}$ & $\mc{S}$ & E1 & E2 & E3 & E4 & E5 & E6 & R1 & R2 & R3 & R4 & R5
\\
\hline
$\mc{M}_{010}$ & $\color{green}\checkmark$ & $\color{green}\checkmark$ & $\color{green}\checkmark$ & $\color{green}\checkmark$ & $\color{green}\checkmark$ & $\color{green}\checkmark$ & $\color{green}\checkmark$ & $\color{green}\checkmark$ & $\color{green}\checkmark$ & $\color{green}\checkmark$ &
   $\color{green}\checkmark$ & $\color{green}\checkmark$ & $\color{green}\checkmark$ & $\color{green}\checkmark$ \\
\hline
$\mc{M}_{020}$ &  $\color{green}\checkmark$ & $\color{green}\checkmark$ & $\color{green}\checkmark$ & $\color{red}\times$ & $\color{green}\checkmark$ & $\color{red}\times$ & $\color{red}\times$ & $\color{green}\checkmark$ & $\color{green}\checkmark$ &
   $\color{red}\times$ & $\color{green}\checkmark$ & $\color{red}\times$ & $\color{green}\checkmark$ & $\color{red}\times$ \\
\hline
\end{tabular}
\end{center}

We address two families of mass terms which are off-diagonal in valley space.
The first family consists of $\mc{M}_{010}$ and $\mc{M}_{020}$, which are local, spin-independent terms. In graphene, O-type Kekulé distortion of the lattice is known to open a gap \cite{chamon2000solitons}. Its periodicity is given by the reciprocal of $\bd{G} = \bd{K}_{D1} - \bd{K}_{D4}$, resulting in valley-mixing bilinear terms.
In kagome, nevertheless, hopping amplitude anisotropy due to such distortion is not responsible for gap-opening. 
However, when the coupling within the rings is strong, a periodic charge ordering emerges. 
This is known as the “plaquette ordered phase", pictured in Figure \ref{fig:plaquettes_kagome}.
To first order, the effect of such background ordering is described by a mean-field potential of the form:
\begin{equation}
    V(r) = \text{Re}\l(\xi \r) \cos \l(\bd{r} \cdot \bd{G} \r)+
    \text{Im}(\xi) \sin \l(\bd{r} \cdot \bd{G} \r)
\end{equation}
whose Fourier transform reads
\begin{equation}
\begin{split}
    \delta H 
    = \frac{\text{Re}(\xi)}{2} \sum_{\bd{k}}
    \l[
    \hat{A}^\dg_{\bd{k}} 
    \hat{A}_{\bd{k}-\bd{G}}
    +
    \hat{A}^\dg_{\bd{k}} 
    \hat{A}_{\bd{k}+\bd{G}}
    \r] +
    \\
     +
    \frac{\text{Im}(\xi)}{2i} \sum_{\bd{k}}
    \l[
    \hat{A}^\dg_{\bd{k}} 
    \hat{A}_{\bd{k}-\bd{G}}
    -
    \hat{A}^\dg_{\bd{k}} 
    \hat{A}_{\bd{k}+\bd{G}}
    \r] + 
    \\
    +
    \boxed{\hat{A} \leftrightarrow \hat{B}} +
    \boxed{\hat{A} \leftrightarrow \hat{C}}
\end{split}
\end{equation}
At the Dirac points these terms are proportional to $\mc{M}_{010}$ and $\mc{M}_{020}$.
All the quantum symmetries
(time-reversal, charge, chiral symmetry) are preserved. 

\begin{center}
\tabcolsep=0.06cm
\small
\begin{tabular}{|c|c|c|c|c|c|c|c|c|c|c|c|c|c|c|}
\hline
  & $\mc{T}$ & $\mc{C}$ & $\mc{S}$ & E1 & E2 & E3 & E4 & E5 & E6 & R1 & R2 & R3 & R4 & R5
\\
\hline
$\mc{M}_{110}$ & $\color{red}\times$ & $\color{red}\times$ & $\color{green}\checkmark$ & $\color{green}\checkmark$ & $\color{red}\times$ & $\color{red}\times$ & $\color{red}\times$ & $\color{red}\times$ &
   $\color{red}\times$ & $\color{red}\times$ & $\color{red}\times$ & $\color{red}\times$ & $\color{red}\times$ & $\color{red}\times$ \\
\hline
$\mc{M}_{210}$  & $\color{red}\times$ & $\color{red}\times$ & $\color{green}\checkmark$ & $\color{red}\times$ & $\color{green}\checkmark$ & $\color{red}\times$ & $\color{red}\times$ & $\color{red}\times$ &
   $\color{red}\times$ & $\color{red}\times$ & $\color{red}\times$ & $\color{red}\times$ & $\color{red}\times$ & $\color{red}\times$ \\
\hline
$\mc{M}_{310}$ & $\color{red}\times$ & $\color{red}\times$ & $\color{green}\checkmark$ & $\color{red}\times$ & $\color{red}\times$ & $\color{red}\times$ & $\color{red}\times$ & $\color{red}\times$ &
   $\color{red}\times$ & $\color{green}\checkmark$ & $\color{green}\checkmark$ & $\color{green}\checkmark$ & $\color{green}\checkmark$ & $\color{green}\checkmark$ \\
\hline
$\mc{M}_{120}$  & $\color{red}\times$ & $\color{red}\times$ & $\color{green}\checkmark$ & $\color{red}\times$ & $\color{red}\times$ & $\color{red}\times$ & $\color{red}\times$ & $\color{red}\times$ &
   $\color{red}\times$ & $\color{red}\times$ & $\color{red}\times$ & $\color{green}\checkmark$ & $\color{red}\times$ & $\color{red}\times$ \\
\hline
$\mc{M}_{220}$ & $\color{red}\times$ & $\color{red}\times$ & $\color{green}\checkmark$ & $\color{green}\checkmark$ & $\color{green}\checkmark$ & $\color{red}\times$ & $\color{red}\times$ & $\color{red}\times$ &
   $\color{red}\times$ & $\color{red}\times$ & $\color{red}\times$ & $\color{green}\checkmark$ & $\color{red}\times$ & $\color{red}\times$ \\
\hline
$\mc{M}_{320}$ & $\color{red}\times$ & $\color{red}\times$ & $\color{green}\checkmark$ & $\color{green}\checkmark$ & $\color{red}\times$ & $\color{green}\checkmark$ & $\color{green}\checkmark$ & $\color{red}\times$ & $\color{red}\times$ &
   $\color{red}\times$ & $\color{green}\checkmark$ & $\color{red}\times$ & $\color{green}\checkmark$ & $\color{red}\times$ \\
\hline
\end{tabular}
\end{center}
The second family resembles the first while showing an additional spin dependence:
$\mc{M}_{j10}$ and $\mc{M}_{j20}$, where $j=1,2,3$.
These terms break time reversal and charge symmetry while preserving chirality.

\begin{figure}[htp]
    \centering
    \includegraphics[width=\linewidth]{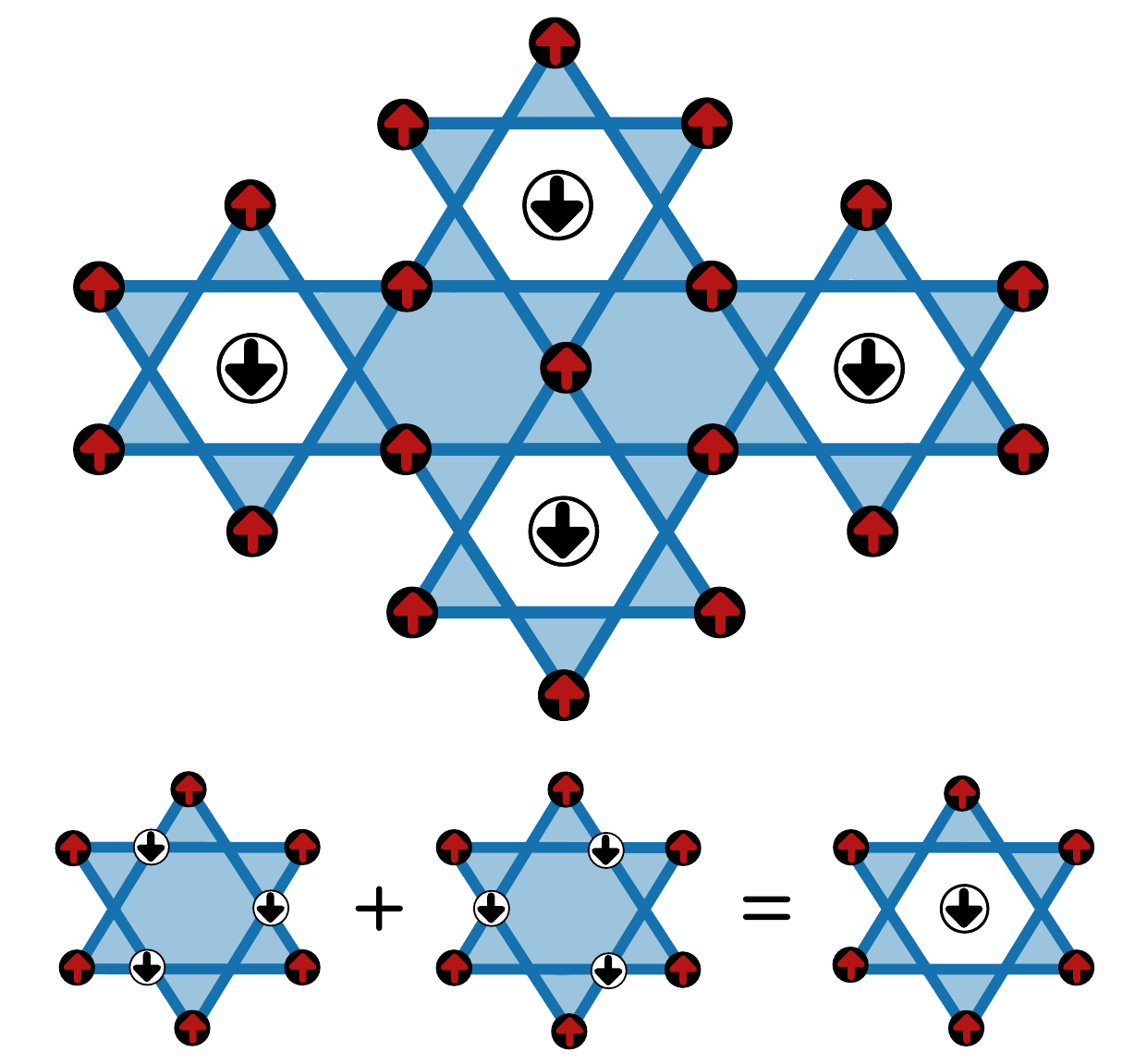}
    \caption{
    \small
    Kagome lattice at 1/3 filling of strongly interacting electrons is in a resonating plaquette phase \cite{pollmann2014interplay}.
    An antiferromagnetic perturbation lifts the spin-degeneracy of the ground state, favouring a magnetic ordering: three resonating, parallel spins form a 3/2 spin on the hexagon, surrounded by localised spins pointing in the opposite direction. A mean-field treatment of the anti-ferromagnetic interaction in the presence of such ordering results in the mass terms $\mc{M}_{j10}$ and $\mc{M}_{j20}$.}
    \label{fig:resonating_plaquette_Kagome}
\end{figure}

Pollmann et al. \cite{pollmann2014interplay} studied a mechanism to spontaneously create these terms. They studied an extended Hubbard model on the kagome lattice which includes on-site and NN repulsion.
In the strong-coupling limit, the effective interaction is decoupled in two terms, accounting for charge and spin exchanges, respectively.
Two distinct spin-ordered phases emerge, depending on the relative strength of the two contributions. In the so-called “short-loop" phase, electrons form loops with shorter possible length, as depicted in Figure\ref{fig:plaquettes_kagome}. A spin-ordering within the rings is also present.
Such phase resembles an O-type Kekulé dimerization pattern with spin dependence, and periodicity given by a wave vector $\bd{G} = \bd{K}_{D1}-\bd{K}_{D4}$, which results in a valley mixing.
In the “plaquette phase", electrons are distributed as sketched in Figure \ref{fig:resonating_plaquette_Kagome}: three electrons resonating on the hexagons are surrounded by three localised ones.
We observe antiferromagnetic spin ordering, with periodicity given by $\bd{G}$.

One way to account for these phenomena in our framework is to treat the Hubbard model in the mean-field approximation. The two effective interactions are then approximated as bilinear terms.
If either short-loop or plaquette phase are realised, one obtains spin-dependent, valley-mixing, on-site potential terms, which correspond to $\mc{M}_{j10}$ and $\mc{M}_{j20}$, with freedom on the spin index. 

\subsection{Dimerization and trimerization}\label{sec:dimerization_kagome}
\begin{figure}[htp]
    \centering
    \includegraphics[width=0.6\linewidth]{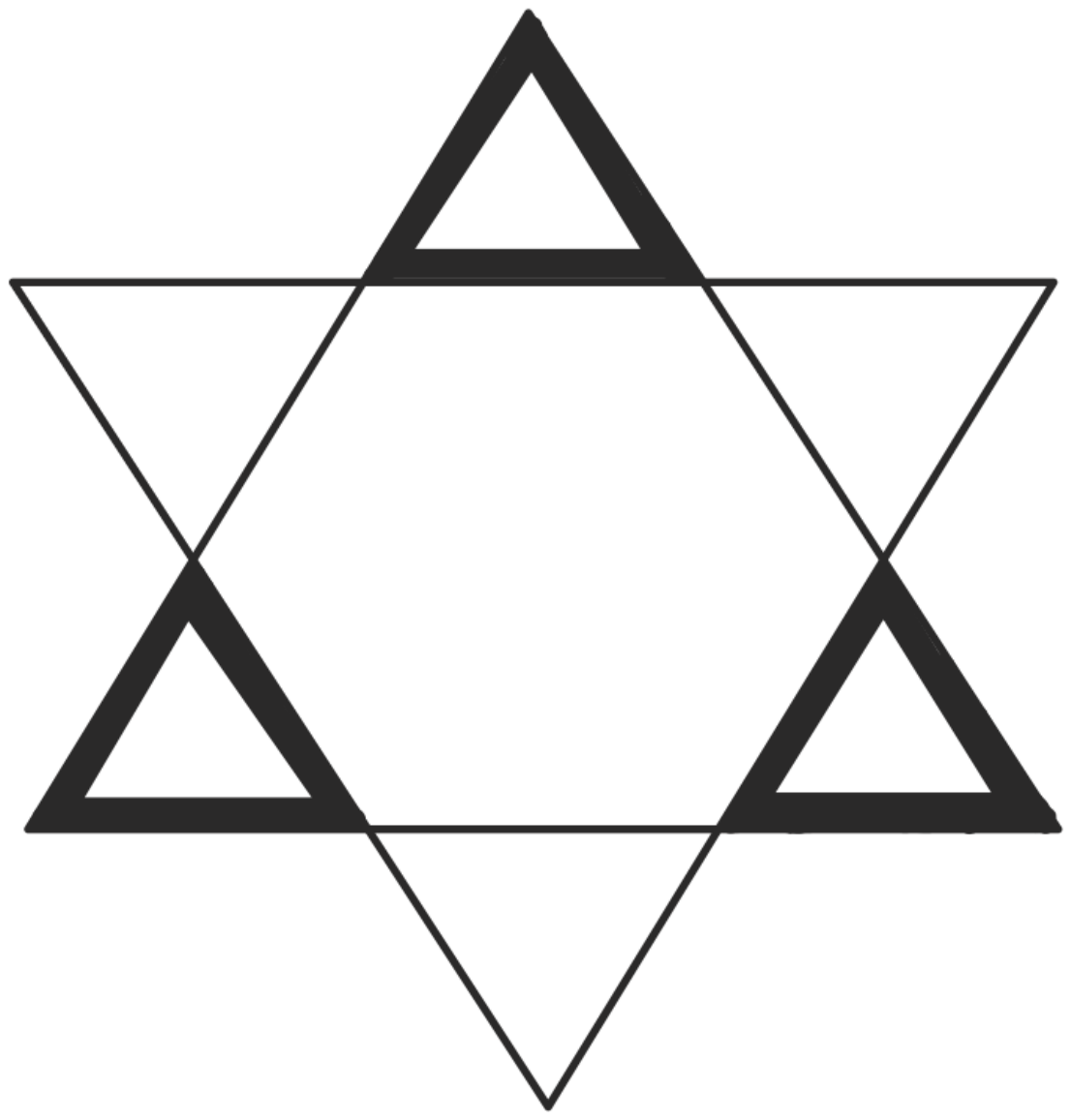}
    \caption{
    \small
    Trimerization pattern of the kagome lattice: thick (thin) lines depict stronger (weaker) bonds. The resulting distortion in hopping amplitude opens a gap at Dirac points, taking the form of the mass term $\mc{M}_{032}$.}
    \label{fig:trimerization_kagome}
\end{figure}

\begin{center}
\tabcolsep=0.06cm
\small
\begin{tabular}{|c|c|c|c|c|c|c|c|c|c|c|c|c|c|c|}
\hline
\multicolumn{1}{|c|}{\multirow{2}{*}{ $\mc{M}_{032}$}}  & $\mc{T}$ & $\mc{C}$ & $\mc{S}$ & E1 & E2 & E3 & E4 & E5 & E6 & R1 & R2 & R3 & R4 & R5
\\
\cline{2-15}
\multicolumn{1}{|c|}{} & $\color{green}\checkmark$ & $\color{red}\times$ & $\color{red}\times$ & $\color{green}\checkmark$ & $\color{red}\times$ & $\color{green}\checkmark$ & $\color{green}\checkmark$ & $\color{red}\times$ & $\color{red}\times$ &
   $\color{red}\times$ & $\color{green}\checkmark$ & $\color{red}\times$ & $\color{green}\checkmark$ & $\color{red}\times$ \\
\hline
\end{tabular}
\end{center}
We consider a trimerized phase\cite{ciola2021chiral}, represented in Figure \ref{fig:trimerization_kagome}.
Such a structure is given by bonds with alternating values of hopping amplitude
$t \to t \pm \eta$, and it is sometimes named “breathing kagome" phase.
The correction to the Bloch Hamiltonian Eq.\ref{eq:BlochHamiltonian} due to trimerization is
\begin{align}
    &h_{\eta}(\bd{k}) = 2 \eta i
    \everymath{\scriptstyle}
    \begin{pmatrix}
    \everymath{\scriptstyle}
    0 & \sin\l(\bd{k}\cdot\bd{\delta}_{AB}\r) & \sin\l(\bd{k}\cdot \bd{\delta}_{AC}\r)
    \\
    -\sin\l(\bd{k}\cdot\bd{\delta}_{AB}\r)   & 0 & \sin\l(\bd{k}\cdot\bd{\delta}_{BC}\r)
    \\
    -\sin\l(\bd{k}\cdot \bd{\delta}_{AC}\r) & -\sin\l(\bd{k}\cdot\bd{\delta}_{BC}\r) & 0
    \end{pmatrix}
\end{align}
with respective field operator $\hat{\Psi}_{\bd{k}} = \Big(\hat{A}_{\bd{k}}, \hat{B}_{\bd{k}}, \hat{C}_{\bd{k}} \Big)^T$.
A second term is added to the original tight-binding Hamiltonian, which can be expanded around Dirac points:
\begin{equation}
\begin{split}
    \mc{H}_{\eta} =
    2 \eta i
    \l[
    s_0 \otimes \sigma_3 \otimes
    \begin{pmatrix}
    0 & 1 & 1
    \\
    -1 & 0 & -1
    \\
    -1 & 1 & 0
    \end{pmatrix}
    \r]
\end{split}
\end{equation}
where the spin and valley degrees of freedom have been included.
This Hamiltonian corresponds to the gap-opening term $\mc{M}_{032}$, which breaks sublattice symmetry while keeping time-reversal intact.
The same gap term is obtained with a direction-dependent dimerization pattern, given by a $\bd{\eta} = \l( \eta_{AB}, \eta_{AC}, \eta_{BC} \r)$ as long as $\bd{\eta} \neq \l(0,0,0\r)$.

There is a correspondence between gap-opening deformations in honeycomb and kagome lattices. 
In graphene, a gap is opened by a staggered, on-site potential, while in kagome by lattice dimerization. 
It similarly occurs for O-type Kekulé distortions: graphene is gapped because of the hopping amplitude distortion, while kagome materials because of the on-site periodic potential. 
This relation is not accidental: kagome lattice at 1/3 filling can be mapped into a dimer model on the honeycomb lattice \cite{pollmann2014interplay}. Quantum states (along with bilinear terms) which are localised on the sites are mapped into states which live on the bonds, and viceversa.
This relation holds equivalently for gapping terms.

\subsection{Spin-orbit coupling in trimerized kagome lattice}\label{sec:SOCtrimerizedKagome}
\begin{figure}[htp]
    \centering
    \includegraphics[width=1.02\linewidth]{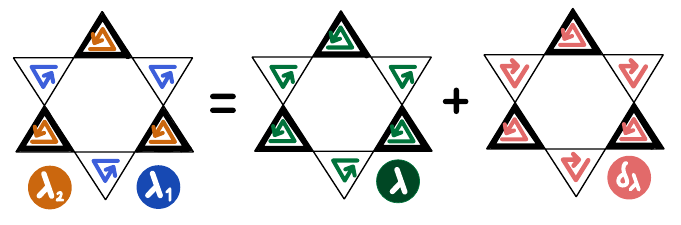}
    \caption{\small Spin-orbit coupling in the trimerized phase, where thick (thin) lines represent strong (weak) bonds. Arrows mark the direction of positive amplitudes hopping for a spin-up. Down spins have reversed arrows.
    The alternating amplitude pattern $\lambda_{1,2}$ (depicted in blue and orange in the LHS) can be decoupled into a
    regular SOC pattern (green arrows) and a contribution from the anisotropy (pink arrows). The first term results in the mass terms $\mc{M}_{j02}$ of Section \ref{sec:SOC_kagome}, while the second in the terms $\mc{M}_{j32}$.}
    \label{fig:SOC_trimerizedKagome}
\end{figure}
\begin{center}
\tabcolsep=0.06cm
\small
\begin{tabular}{|c|c|c|c|c|c|c|c|c|c|c|c|c|c|c|}
\hline
  & $\mc{T}$ & $\mc{C}$ & $\mc{S}$ & E1 & E2 & E3 & E4 & E5 & E6 & R1 & R2 & R3 & R4 & R5
\\
\hline
$\mc{M}_{132}$ & $\color{red}\times$ & $\color{green}\checkmark$ & $\color{red}\times$ & $\color{green}\checkmark$ & $\color{green}\checkmark$ & $\color{red}\times$ & $\color{red}\times$ & $\color{red}\times$ &
   $\color{red}\times$ & $\color{red}\times$ & $\color{red}\times$ & $\color{green}\checkmark$ & $\color{red}\times$ & $\color{red}\times$ \\
\hline
$\mc{M}_{232}$ & $\color{red}\times$ & $\color{green}\checkmark$ & $\color{red}\times$ & $\color{red}\times$ & $\color{red}\times$ & $\color{red}\times$ & $\color{red}\times$ & $\color{red}\times$ &
   $\color{red}\times$ & $\color{red}\times$ & $\color{red}\times$ & $\color{green}\checkmark$ & $\color{red}\times$ & $\color{red}\times$ \\
\hline
$\mc{M}_{332}$ & $\color{red}\times$ & $\color{green}\checkmark$ & $\color{red}\times$ & $\color{red}\times$ & $\color{green}\checkmark$ & $\color{red}\times$ & $\color{red}\times$ & $\color{green}\checkmark$ & $\color{green}\checkmark$ &
   $\color{red}\times$ & $\color{green}\checkmark$ & $\color{red}\times$ & $\color{green}\checkmark$ & $\color{red}\times$ \\
\hline
\end{tabular}
\end{center}
We investigate how the “breathing kagome" trimerisation pattern affects spin-orbit coupling (SOC).
Following the procedure of Section \ref{sec:SOC_kagome} we look at the SOC in this configuration: the lattice anisotropy is translated in two different SOC amplitudes\cite{bolens2019topological} $\pm \lambda_1$ and $\pm \lambda_2$, when they are associated to a stronger or weaker bond respectively \cite{nagaosa2010anomalous}. 
Again, the sign of such amplitudes depends on the hopping direction and the electron spin pointing up (or down).
One can decouple the spin-orbit interaction into a regular SOC pattern with the amplitude given by the average of $\lambda_1$, $\lambda_2$ and a hopping pattern which takes into account the trimerization effects.
\begin{equation}
    \lambda = \frac{\lambda_1 + \lambda_2}{2}
    \qquad 
    \delta\lambda = \frac{\lambda_1 - \lambda_2}{2}
\end{equation}
The first contribution is equal to Equation \ref{eq:SOC_kagome}, and therefore corresponds to the gap term $\mc{M}_{302}$. The second corresponds to
\begin{equation}
\begin{split}
    \delta H_{soc} = 
    i \frac{\delta \lambda}{2} \sum_{i \in A}  
    \Big( 
    \hat{A}^\dg_{i,\u} 
    \hat{B}_{i+\delta_{AB}, \u}
    -
    \hat{A}^\dg_{i,\u} 
    \hat{B}_{i-\delta_{AB}, \u}
    +
    \\
    -
    \boxed{\hat{B} 
    \leftrightarrow
    \hat{C}}
    -
    \boxed{\hat{A}
    \leftrightarrow
    \hat{C}}
    -
    \boxed{\u 
    \hspace{0.2cm}
    \leftrightarrow
    \hspace{0.2cm}\d}
    \Big)
    +
    h.c.
\end{split}
\end{equation}
where we stress that for a fixed spin (up or down) the trimerization results in an alternating pattern of SOC strength.
After Fourier transformation, the Hamiltonian reads
\begin{equation}
\begin{split}
    & \delta  \mc{H}_{soc}(\bd{k}) =
    \sum_{\bd{k}}
    \hat{\Phi}^\dg_{\bd{k}}
    \l[ 
    s_3 \otimes
    \delta h_{soc}(\bd{k})
    \r]
    \hat{\Phi}_{\bd{k}}
    \\
    & \delta h(\bd{k})_{soc} = 
    \everymath{\scriptstyle}
    i \delta \lambda 
    \begin{pmatrix}
    0 & \sin\l(\bd{k}\cdot\bd{\delta}_{AB}\r) & -\sin\l(\bd{k}\cdot\bd{\delta}_{AC}\r)
    \\
    -\sin\l(\bd{k}\cdot\bd{\delta}_{AB}\r) & 0 & \sin\l(\bd{k}\cdot\bd{\delta}_{BC}\r)
    \\
    \sin\l(\bd{k}\cdot\bd{\delta}_{AC}\r) & -\sin\l(\bd{k}\cdot\bd{\delta}_{BC}\r)  & 0
    \end{pmatrix}
\end{split}
\end{equation}
This Hamiltonian is odd in $\bd{k}$: the expansion around the two Dirac points results in opposite signs.
\begin{equation}
    \delta \mc{H}_{soc} = 
    i \frac{\delta \lambda}{2}
    \hspace{0.07cm}
    s_3 \otimes
    \sigma_3 \otimes
    \begin{pmatrix}
    0 & 1 & -1
    \\
    -1 & 0 & -1
    \\
    1 & 1 & 0
    \end{pmatrix}
\end{equation}
which corresponds to the gap-opening term $\mc{M}_{332}$.
As one can see from Table \ref{tab:gaps_kagome} and Figure \ref{fig:SOC_trimerizedKagome} (second contribution, pattern of arrows), such term is invariant under rotations of $\pm \frac{2\pi}{3}$ and under certain reflections. 
When the effect of trimerization is taken into account both for hopping amplitude and spin-orbit coupling ($\mc{M}_{032}$ and $\mc{M}_{332}$ respectively), all reflection symmetries are broken. 
Again, a non-planar setup gives rise to $x$ and $y$ components in the spin-coupling, which correspond to the gap-opening terms 
$\mc{M}_{132}$ and $\mc{M}_{232}$.

\section{\label{sec:conclusions} Conclusions}
Motivated by recent attention on kagome materials we investigated the properties of free electrons on such structures when their filling is $1/3$.
In this case, a Dirac theory of massless quasiparticles emerges, like the one in graphene.
We derived a family of sixteen bilinear Hamiltonians which open a gap at the Dirac points. When such terms are present the quasiparticles gain mass, and the system turns into an insulator.
We proceed to identify microscopic realizations of said gap terms, either through explicit perturbations or spontaneously originating from symmetry breaking. 
A similar discussion was present in the literature on graphene, but not similarly structured for kagome materials. In the latter, however, especially interaction-related effects could be enhanced due to the boosted fine-structure constant.
Our main findings can be summarized as follows:
The anomalous Hall effect (Haldane's model) and the quantum spin-Hall effect (spin-orbit coupling), originally proposed in graphene, exist in kagome materials with similar consequences, possibly with more favourable conditions:
contrarily to graphene, these effects emerge on the kagome lattice already for nearest-neighbour hoppings, which indicates the latter as a prominent host for these phenomena.
A different category of instabilities concerns lattice deformation.
The kagome lattice at 1/3 filling can be mapped into a dimer model on the honeycomb lattice of graphene; remarkably, this relation emerges explicitly when comparing graphene and kagome gaps: terms living on the lattice sites of graphene have a correspondent living on the bond of kagome materials, and vice versa.
The simplest, gap-opening lattice deformation is represented by a dimerization of the kagome lattice, which corresponds to the staggered potential in graphene. 
Moreover, we considered distortion patterns with enlarged unit cells that couple the two valleys in momentum space, such as O-type Kekulé patterns. Again, such patterns open a gap in the two materials in two conjugate ways: Graphene quasiparticles become massive because of the hopping amplitude distortion, while in kagome as on-site potential with identical periodicity.
Lastly, we studied masses originating from antiferromagnetic interaction.
Honeycomb lattice at 1/2 filling is a perfect host for Néel ordering, and a mean-field treatment reveals the trivial gapped nature of the antiferromagnetic phase.
In kagome, the corner-sharing triangles frustrate antiferromagnetic orderings. 
In this work, however, we focus on 1/3 filling of the kagome lattice, with, on average, 2 electrons for each triangle. The presence of empty sites leaves space for different antiferromagnetic orderings. We explored the resonating plaquette phase, characterised by enlarged magnetic cells, and known to emerge in the strong-coupling limit. Similarly to graphene, a mean-field treatment reveals the gapped nature of this phase

\section*{Acknowledgements}
The authors want to thank Riccardo Ciola and Ronny Thomale for discussions and collaborations on related subjects.

\appendix
\section{\label{sec:appendix}Lattice symmetries}
Both honeycomb and kagome lattice symmetries are encoded in the wallpaper group p6m: it consists of 6 reflection axis, 1 rotation centre of order six, 2 of order three, and 3 of order two \cite{flores2007classifying}.
In this appendix, we will derive the operator for these geometrical transformations, and study their relation with the Hamiltonian and the mass terms.
In the low-energy approximation, the symmetry relations require additional care: one has to consider the starting point of the expansion and how it transforms under the symmetry. E.g., 
$E_1$ transforms $\bd{K}_{D1}$ into $\bd{K}_{D4}$; $E_2$ instead keeps $\bd{K}_{D1}$ unchanged (See Fig. \ref{fig:reflections}). 
In general, these operations transform the Hamiltonian expanded around two Dirac points - let's say $\bd{K}_{Di},\bd{K}_{Dj}$ - into one Hamiltonian expanded around two different Dirac points $\bd{K}_{Dk},\bd{K}_{Dl}$. 
While the full Hamiltonian enjoys all the symmetry of the model (any of these transformations sends $H\to H$), its low-energy approximations are \textit{less symmetric}, meaning that any of these might be transformed one into another under symmetry operations: $H_{ij} \to H_{kl}$. 

\subsubsection{Gauge transformation} \label{sec:gauge_transformation}
Before deriving the representation of the lattice symmetry transformations, we emphasize that extra care must be taken in the classification of the mass terms according to these symmetries.
Rotations and reflections operators have in fact a natural representation in the original full Hilbert space $\mathbb{H}_{\text{full}}$. 
Therefore, it is convenient to embed the gap terms in $\mathbb{H}_{\text{full}}$ in order to to classify them.
As mentioned in Sec. \ref{sec:massterms}, this is achieved by following the prescription of Eq. (\ref{embed_mass_term_in_H_full}) and subsequently applying the inverse of the transformation (dependent on the Dirac point choice) originally used in the low energy expansion to decouple the flat band from the emerging Dirac theory. 
However, given a certain choice for the Dirac points, the corresponding mass terms embedded in the full Hilbert space do not automatically open a gap in the spectrum of the Bloch Hamiltonian expanded around a different pair of valleys. In fact, in order to obtain the corresponding mass terms in $\mathbb{H}_{\text{full}}$ for a different set of valleys, an additional \textit{gauge transformation} has to be employed.
This is more effectively explained with an example:
In Sec. \ref{sec:massterms} we have expanded $h( \boldsymbol{k} )$ around $\bd{K}_{D1}, \bd{K}_{D4}$, and derived the emerging Dirac theory with its corresponding mass terms.  
Let us focus, for instance, on the mass term $\mathcal{M}_{002}$, and its representation in $\mathbb{H}_{\text{full}}$, denoted by $M^{14}_{002}$ (here, the upper indices indicate the valleys choice).
Given the equivalence between Dirac points in the first Brillouin zone, we expect $M^{14}_{002}$ to open a gap at all the valleys.  
However, this turns out to not be the case, and, for instance, $H_{32}+M_{002}^{14}$ remains gap-less.
The reason for this seeming contradiction is that the equivalence between Dirac points is not apparent in the form of $h( \boldsymbol{k} )$. 
In fact, upon evaluating it at equivalent Dirac points, say $\bd{K}_{D1}, \bd{K}_{D3}$, we obtain two different representations:
\begin{equation}
    h(\bd{K}_{D1})
    \everymath{\scriptstyle}
    =
    \begin{pmatrix}
      0 & -1 & 1\\
      -1 & 0 & 1\\
      1 & 1 & 0
    \end{pmatrix} \text{,} \quad
    h(\bd{K}_{D3})
    \everymath{\scriptstyle}
    =
    \begin{pmatrix}
      0 & 1 & 1\\
      1 & 0 & -1\\
      1 & -1 & 0
    \end{pmatrix}
\end{equation}
The equivalence between certain valleys is however still encoded in the fact that $h(\bd{K}_{D1})$ and $h(\bd{K}_{D3})$ exhibit the same spectrum. 
This suggests that there exists a gauge transformation connecting them. In this example, it is given by
\begin{equation}
    g_{13}=\begin{pmatrix}
      1 & 0 & 0\\
      0 & -1 & 0\\
      0 & 0 & 1
    \end{pmatrix},
\end{equation}
such that 
\begin{equation}
    g_{13}\cdot h(\bd{K}_{D1}) \cdot  g_{13}^{-1}=h(\bd{K}_{D3}).
\end{equation}
Similarly, all the other transformations \(g_{ac}\) that connect any pair of Bloch Hamiltonians, \(h(\bd{K}_{Da})\) and \(h(\bd{K}_{Dc})\), evaluated at equivalent valleys $a,c$ can be determined.
The issue presented above ($H_{32}+M_{002}^{14}$ being gap-less) is therefore solved by appropriately gauge-transforming the term $M_{002}^{14}$ to the valleys 3, 2:
\begin{equation}\label{M_002_32}
\everymath{\scriptstyle}
M_{002}^{32} = 
\begin{pmatrix}
     g_{13} & \bd{0} & \bd{0} & \bd{0} \\
     \bd{0} & g_{42} & \bd{0} & \bd{0} \\
     \bd{0}  & \bd{0}  & g_{13} & \bd{0}  \\
     \bd{0}  & \bd{0}  & \bd{0}  & g_{42}
\end{pmatrix}
\cdot M_{002}^{14} \cdot
\everymath{\scriptstyle}
\begin{pmatrix}
     g_{13}^{-1} & \bd{0} & \bd{0}  & \bd{0}  \\
     \bd{0}  & g_{42}^{-1} &\bd{0}  & \bd{0}  \\
     \bd{0}  & \bd{0}  & g_{13}^{-1} & \bd{0}  \\
     \bd{0}  & \bd{0}  & \bd{0}  & g_{42}^{-1}
\end{pmatrix}
\end{equation}
which opens a gap in the spectrum of $H_{32}$ (in Eq. (\ref{M_002_32}), $\bd{0}$ indicates the $3 \times 3$ null matrix).
In fact, this is true in general for all the mass terms, provided that we had appropriately chosen the $U$ transformations to decouple the flat band depending on the Dirac point choice\footnote{The specific form of the low energy theory Eq.\eqref{eq:H2dim} can be derived starting from any choice two of inequivalent Dirac points $\bd{K}_{Da}$, $\bd{K}_{Db}$. This is done by appropriately absorbing these gauge transformations into the flat-band decoupling matrices $U_{ab}$, which therefore depend on the valley choice $a,b$. This guarantees the same family of mass terms $\mc{M}_{ijk}$ independently of the chosen valleys.}.
Schematically:

\bigskip

\begin{tikzpicture}[auto, thick, node distance=2cm, >=Stealth]
    \node (A) at (0, 0) {$\highlight{M_{ijk}^{ab}}$};
    \node (B) at (4, 0) {$\highlight{M_{ijk}^{cd}}$};
    \node (C) at (4, -4) {$\highlight{\mathcal{M}_{ijk}}$};
    \node (D) at (0, -4) {$\highlight{\mathcal{M}_{ijk}}$};

    \node[left=0.5cm of D] {($\mathbb{H}_{\text{eff}}$)};

    \node[left=0.5cm of A] {($\mathbb{H}_{\text{full}}$)};
    \node[above=0.1cm of A] {valleys $a,b$};
    \node[above=0.1cm of B] {valleys $c,d$};

    \draw[<->] (A) -- (B) node[midway, below] {gauge};
    \draw[<->] (B) -- (C) node[midway, right] {embedding};
    \draw[<->] (D) -- (A) node[midway, left] {embedding};
   
\end{tikzpicture}

\subsubsection{Reflections}\label{sec:reflections}
The general procedure to perform reflections consists of two operations: 
a parity transformation (inversion), and a rotation of $\pi$ around one axis $\bd{n}$ perpendicular to the reflection axis.
The spin $\bd{s}$ is invariant under parity and transforms as a spin under the rotation around $\bd{n}$:
\begin{equation}
\begin{split}
    R_{\bd{n}}(\theta) &= \exp \l( - i \theta \bd{n}\cdot\bd{s} \r)
    \\
    &=
    \mathbb{1}  \cos \l(\frac{\theta}{2} \r) - i  \bd{n}\cdot\bd{s} \sin \l(\frac{\theta}{2}\r)
\end{split}
\end{equation}

We will explicitly derive one reflection and one rotation operator. The procedure for the others is equivalent, and the result is reported in Table \ref{tab:symmetry_operators}.

$E1$ reflects with respect to the $y$-axis; it transforms the spin with a rotation of $\pi$ around the $x$-axis: $R_{\bd{x}}(\pi) = -i s_1$.
Its action on the momenta sends $\bd{k} = \l( k_x, k_y \r)  \to \l( -k_x, k_y \r)$. The valleys in the Brillouin zone are then exchanged.
Sublattices transform as 
$A \to B $, $B \to A$, $C \to C$. 
The quantum operator in the Hilbert space $\mathbb{H_{\text{full}}}$ is then
\begin{align}
    E_1 
    &= -i s_1 \otimes \sigma_1 \otimes 
    \begin{pmatrix}
        0 & 1 & 0 \\
        1 & 0 & 0 \\
        0 & 0 & 1 \\
    \end{pmatrix}\;.
\end{align}
The low-energy Hamiltonian transforms as
\begin{align*}
    \qquad \qquad E_1 H_{14}(k_x, k_y) E_1^{-1} = H_{36}(-k_x, k_y)\;.
\end{align*}

\subsubsection{Rotations}
In this section, we consider rotation around the centre of the hexagons (centre of order six).

R1 is an anti-clockwise rotation of $\pi/3$. It exchanges the kagome sublattices as $A \to C$, $B \to A$, $C \to B$. Its action on the momenta transforms $\bd{k} = \l( k_x, k_y \r)  \to  \bd{k}'  = \l(\frac{k_x}{2} + \frac{\sqrt{3}}{2}k_y, -\frac{\sqrt{3}}{2}k_x + \frac{k_y}{2}\r)$.
The full operator is given by
\begin{align}
    &R_{1} = \frac{1}{2} \l(\sqrt{3} s_0 -
    i s_3 \r) \otimes \sigma_1 \otimes 
    \begin{pmatrix}
     0 & 0 & 1 \\
     1 & 0 & 0 \\
     0 & 1 & 0 \\
    \end{pmatrix}\;.
\end{align}
The Hamiltonian transforms according to
\begin{align*}
    R_{1} H_{14}(\bd{k}) R_{1}^{-1} = H_{25}(\bd{k}')\;.
\end{align*}

\begin{table*}[b]
\small
\centering
\begin{tabular}{|c||c|c|c|c||c|c|c|c|c|c||c|c|c|c|c|}
\hline
\multicolumn{16}{c}{\textbf{Symmetries of the mass terms $\mc{M}_{ijk}$}} \\
\hline        
Idx $\{ijk\}$ & Parity & T-rev & Char & Chiral & E1 & E2 & E3 & E4 & E5 & E6 & R1 & R2 & R3 & R4 & R5 \\ 
         \hline\hline
 \{0,0,2\} & $\color{green}\checkmark$ & $\color{red}\times$ & $\color{green}\checkmark$ & $\color{red}\times$ & $\color{red}\times$ & $\color{red}\times$ & $\color{red}\times$ & $\color{red}\times$ & $\color{red}\times$ &
   $\color{red}\times$ & $\color{green}\checkmark$ & $\color{green}\checkmark$ & $\color{green}\checkmark$ & $\color{green}\checkmark$ & $\color{green}\checkmark$ \\
 \{0,1,0\} & $\color{green}\checkmark$ & $\color{green}\checkmark$ & $\color{green}\checkmark$ & $\color{green}\checkmark$ & $\color{green}\checkmark$ & $\color{green}\checkmark$ & $\color{green}\checkmark$ & $\color{green}\checkmark$ & $\color{green}\checkmark$ & $\color{green}\checkmark$ & $\color{green}\checkmark$ &
   $\color{green}\checkmark$ & $\color{green}\checkmark$ & $\color{green}\checkmark$ & $\color{green}\checkmark$ \\
 \{0,2,0\} & $\color{red}\times$ & $\color{green}\checkmark$ & $\color{green}\checkmark$ & $\color{green}\checkmark$ & $\color{red}\times$ & $\color{green}\checkmark$ & $\color{red}\times$ & $\color{red}\times$ & $\color{green}\checkmark$ & $\color{green}\checkmark$ &
   $\color{red}\times$ & $\color{green}\checkmark$ & $\color{red}\times$ & $\color{green}\checkmark$ & $\color{red}\times$ \\
 \{0,3,2\} & $\color{red}\times$ & $\color{green}\checkmark$ & $\color{red}\times$ & $\color{red}\times$ & $\color{green}\checkmark$ & $\color{red}\times$ & $\color{green}\checkmark$ & $\color{green}\checkmark$ & $\color{red}\times$ & $\color{red}\times$ &
   $\color{red}\times$ & $\color{green}\checkmark$ & $\color{red}\times$ & $\color{green}\checkmark$ & $\color{red}\times$ \\
\hline
 \{1,1,0\} & $\color{green}\checkmark$ & $\color{red}\times$ & $\color{red}\times$ & $\color{green}\checkmark$ & $\color{green}\checkmark$ & $\color{red}\times$ & $\color{red}\times$ & $\color{red}\times$ & $\color{red}\times$ &
   $\color{red}\times$ & $\color{red}\times$ & $\color{red}\times$ & $\color{red}\times$ & $\color{red}\times$ & $\color{red}\times$ \\
 \{2,1,0\} & $\color{green}\checkmark$ & $\color{red}\times$ & $\color{red}\times$ & $\color{green}\checkmark$ & $\color{red}\times$ & $\color{green}\checkmark$ & $\color{red}\times$ & $\color{red}\times$ & $\color{red}\times$ &
   $\color{red}\times$ & $\color{red}\times$ & $\color{red}\times$ & $\color{red}\times$ & $\color{red}\times$ & $\color{red}\times$ \\
 \{3,1,0\} & $\color{green}\checkmark$ & $\color{red}\times$ & $\color{red}\times$ & $\color{green}\checkmark$ & $\color{red}\times$ & $\color{red}\times$ & $\color{red}\times$ & $\color{red}\times$ & $\color{red}\times$ &
   $\color{red}\times$ & $\color{green}\checkmark$ & $\color{green}\checkmark$ & $\color{green}\checkmark$ & $\color{green}\checkmark$ & $\color{green}\checkmark$ \\
 \{1,2,0\} & $\color{red}\times$ & $\color{red}\times$ & $\color{red}\times$ & $\color{green}\checkmark$ & $\color{red}\times$ & $\color{red}\times$ & $\color{red}\times$ & $\color{red}\times$ & $\color{red}\times$ &
   $\color{red}\times$ & $\color{red}\times$ & $\color{red}\times$ & $\color{green}\checkmark$ & $\color{red}\times$ & $\color{red}\times$ \\
 \{2,2,0\} & $\color{red}\times$ & $\color{red}\times$ & $\color{red}\times$ & $\color{green}\checkmark$ & $\color{green}\checkmark$ & $\color{green}\checkmark$ & $\color{red}\times$ & $\color{red}\times$ & $\color{red}\times$ &
   $\color{red}\times$ & $\color{red}\times$ & $\color{red}\times$ & $\color{green}\checkmark$ & $\color{red}\times$ & $\color{red}\times$ \\
 \{3,2,0\} & $\color{red}\times$ & $\color{red}\times$ & $\color{red}\times$ & $\color{green}\checkmark$ & $\color{green}\checkmark$ & $\color{red}\times$ & $\color{green}\checkmark$ & $\color{green}\checkmark$ & $\color{red}\times$ & $\color{red}\times$ &
   $\color{red}\times$ & $\color{green}\checkmark$ & $\color{red}\times$ & $\color{green}\checkmark$ & $\color{red}\times$ \\
\hline
 \{1,0,2\} & $\color{green}\checkmark$ & $\color{green}\checkmark$ & $\color{red}\times$ & $\color{red}\times$ & $\color{red}\times$ & $\color{green}\checkmark$ & $\color{red}\times$ & $\color{red}\times$ & $\color{red}\times$ &
   $\color{red}\times$ & $\color{red}\times$ & $\color{red}\times$ & $\color{red}\times$ & $\color{red}\times$ & $\color{red}\times$ \\
 \{2,0,2\} & $\color{green}\checkmark$ & $\color{green}\checkmark$ & $\color{red}\times$ & $\color{red}\times$ & $\color{green}\checkmark$ & $\color{red}\times$ & $\color{red}\times$ & $\color{red}\times$ & $\color{red}\times$ &
   $\color{red}\times$ & $\color{red}\times$ & $\color{red}\times$ & $\color{red}\times$ & $\color{red}\times$ & $\color{red}\times$ \\
 \{3,0,2\} & $\color{green}\checkmark$ & $\color{green}\checkmark$ & $\color{red}\times$ & $\color{red}\times$ & $\color{green}\checkmark$ & $\color{green}\checkmark$ & $\color{green}\checkmark$ & $\color{green}\checkmark$ & $\color{green}\checkmark$ & $\color{green}\checkmark$ & $\color{green}\checkmark$
   & $\color{green}\checkmark$ & $\color{green}\checkmark$ & $\color{green}\checkmark$ & $\color{green}\checkmark$ \\
\hline
 \{1,3,2\} & $\color{red}\times$ & $\color{red}\times$ & $\color{green}\checkmark$ & $\color{red}\times$ & $\color{green}\checkmark$ & $\color{green}\checkmark$ & $\color{red}\times$ & $\color{red}\times$ & $\color{red}\times$ &
   $\color{red}\times$ & $\color{red}\times$ & $\color{red}\times$ & $\color{green}\checkmark$ & $\color{red}\times$ & $\color{red}\times$ \\
 \{2,3,2\} & $\color{red}\times$ & $\color{red}\times$ & $\color{green}\checkmark$ & $\color{red}\times$ & $\color{red}\times$ & $\color{red}\times$ & $\color{red}\times$ & $\color{red}\times$ & $\color{red}\times$ &
   $\color{red}\times$ & $\color{red}\times$ & $\color{red}\times$ & $\color{green}\checkmark$ & $\color{red}\times$ & $\color{red}\times$ \\
 \{3,3,2\} & $\color{red}\times$ & $\color{red}\times$ & $\color{green}\checkmark$ & $\color{red}\times$ & $\color{red}\times$ & $\color{green}\checkmark$ & $\color{red}\times$ & $\color{red}\times$ & $\color{green}\checkmark$ & $\color{green}\checkmark$ &
   $\color{red}\times$ & $\color{green}\checkmark$ & $\color{red}\times$ & $\color{green}\checkmark$ & $\color{red}\times$ \\
\hline
\end{tabular}
    \caption{\textbf{Kagome gap terms} and their symmetries. The terms are identified
    by the indices $i,j,k$, according to the representation 
    $
    m 
    \mc{M}_{ijk} 
    = m \hspace{0.11cm} 
    s_i \otimes \sigma_j \otimes \tau_k,
    $ 
    where $s$, $\sigma$ and $\tau$ are Pauli matrices respectively acting on the spin, valley and $\textit{sublattice}$ degree of freedom (for the definition of the latter see main text).
    $E_{i}$ and $R_{j}$ are respectively the reflection and rotation operators as defined in Fig. (\ref{fig:reflections}).}
    \label{tab:gaps_kagome}
\end{table*}

\begin{table*}[h]
    \centering
    \begin{tabular}{||c|c||}
    \hline
    \hline
    \rule{0pt}{6ex}    
    $E_1$ & $-i s_1 \otimes \sigma_1 \otimes 
    \begin{pmatrix}
        0 & 1 & 0 \\
        1 & 0 & 0 \\
        0 & 0 & 1 \\
    \end{pmatrix}$
    \\[0.55cm]
    \hline
    \rule{0pt}{6ex}  
    $E_2$ & $-\frac{i}{2} \l( -s_1 + \sqrt{3} s_y \r)\otimes \sigma_1 \otimes 
    \begin{pmatrix}
        1 & 0 & 0 \\
        0 & 0 & 1 \\
        0 & 1 & 0 \\
    \end{pmatrix}$
    \\[0.55cm]
    \hline
    \rule{0pt}{6ex}  
    $E_3$ & $-\frac{i}{2} \l( -s_1 + \sqrt{3} s_y \r)\otimes \sigma_1 \otimes 
    \begin{pmatrix}
        1 & 0 & 0 \\
        0 & 0 & 1 \\
        0 & 1 & 0 \\
    \end{pmatrix}$
    \\[0.55cm]
    \hline
    \rule{0pt}{6ex}  
    $E_4$ & $-\frac{i}{2} \l( s_1 + \sqrt{3} s_y \r)\otimes \sigma_1 \otimes 
    \begin{pmatrix}
        0 & 0 & 1 \\
        0 & 1 & 0 \\
        1 & 0 & 0 \\
    \end{pmatrix}$
    \\[0.55cm]
    \hline
    \rule{0pt}{6ex}  
    $E_5$ & $-\frac{i}{2} \l(-\sqrt{3} s_1 +  s_y \r)\otimes \sigma_0 \otimes 
    \begin{pmatrix}
        0 & 0 & 1 \\
        0 & 1 & 0 \\
        1 & 0 & 0 \\
    \end{pmatrix}$
    \\[0.55cm]
    \hline
    \rule{0pt}{6ex}  
    $E_6$ & $-\frac{i}{2} \l(\sqrt{3} s_1 +  s_y \r)\otimes \sigma_0 \otimes 
    \begin{pmatrix}
        1 & 0 & 0 \\
        0 & 0 & 1 \\
        0 & 1 & 0 \\
    \end{pmatrix}$
    \\[0.55cm]
    \hline
    \hline
    \end{tabular}
    \qquad
    \begin{tabular}{||c|c||}
    \hline
    \hline
    \rule{0pt}{6ex}    
    $R_{1}$ & 
    $\frac{1}{2} \l(\sqrt{3} s_0 -
    i s_3 \r) \otimes \sigma_1 \otimes 
    \begin{pmatrix}
     0 & 0 & 1 \\
     1 & 0 & 0 \\
     0 & 1 & 0 \\
    \end{pmatrix}$
    \\[0.55cm]
    \hline
    \rule{0pt}{6ex}  
    $R_{2}$ & 
    $\frac{1}{2} \l(s_0 - i \sqrt{3} s_3 \r) \otimes \sigma_0 \otimes
    \begin{pmatrix}
     0 & 1 & 0 \\
     0 & 0 & 1 \\
     1 & 0 & 0 \\
    \end{pmatrix} $
    \\[0.55cm]
    \hline
    \rule{0pt}{6ex}  
    $R_{3}$ & 
    $ -i s_3 \otimes \sigma_1 \otimes \mathbb{1}_3$
    \\[0.55cm]
    \hline
    \rule{0pt}{6ex}  
    $R_{4}$ & 
    $\frac{1}{2} \l(s_0 + i \sqrt{3} s_3 \r) \otimes \sigma_0 \otimes 
    \begin{pmatrix}
     0 & 0 & 1 \\
     1 & 0 & 0 \\
     0 & 1 & 0 \\
    \end{pmatrix} $
    \\[0.55cm]
    \hline
    \rule{0pt}{6ex}  
    $R_{5}$ & 
    $\frac{1}{2}  \l(\sqrt{3} s_0 + i s_3 \r)  \otimes \sigma_1 \otimes 
     \begin{pmatrix}
     0 & 1 & 0 \\
     0 & 0 & 1 \\
     1 & 0 & 0 \\
    \end{pmatrix} $
    \\[0.55cm]
    \hline
    \hline
    \end{tabular}
    \caption{Reflection and rotation operators. $s$, $\sigma$ are Pauli matrices acting on the spin and valley degree of freedom, while the third component of the tensor product acts in the actual sublattice space. }
    \label{tab:symmetry_operators}
\end{table*}

\bibliography{bibliography}
\nocite{*}

\end{document}